\documentclass[twocolumn,aps]{revtex4-1}
\usepackage{userdefs}

\begin{document}
  \bibliographystyle{naturemag}

% \title{Frustration-driven dimerization in two-leg Heisenberg ladder with trimer rungs}
\title{Dimerization and spin-decoupling in
two-leg Heisenberg ladder with frustrated trimer rungs}

\author{Andreas Weichselbaum}
\email{weichselbaum@bnl.gov}
\author{Weiguo Yin}
\author{Alexei M. Tsvelik}

\affiliation{Condensed Matter Physics and Materials Science Division,
Brookhaven National Laboratory, Upton, New York 11973, USA}

\begin{abstract}

We study the antiferromagnetic spin-half Heisenberg
ladder in the presence of an additional frustrating
rung spin which is motivated and relevant also for the
description of real two-dimensional materials such as
the two-dimensional trimer magnet
Ba$_4$Ir$_3$O$_{10}$. We study the zero-temperature
phase diagram, where we combine numerical and
analytical methods into an overall consistent
description. All numerical simulations are also
accompanied by studies of the dynamical spin structure
factor obtained via the density matrix renormalization
group. Overall, we find in the regime of strong rung
coupling a gapped dimerized phase related to competing
symmetry sectors in Hilbert space that ultimately
results in frustration-driven spin-Peierls transition.
In the weak rung-coupling regime, the system is
uniform, yet shows a gapped spinon continuum together
with a sharp coherent low-energy branch which renders
the system critical overall. In either case, the
additional rung spin quickly get sidelined and nearly
decouple once their bare coupling to the ladder
rops somewhat below the direct Heisenberg
coupling of the legs.

\end{abstract}

\date{\today}

\maketitle

\section{Introduction}

\begin{figure}[b!]
\begin{center}
% \vspace{-.15in}
\includegraphics[width=1\columnwidth]{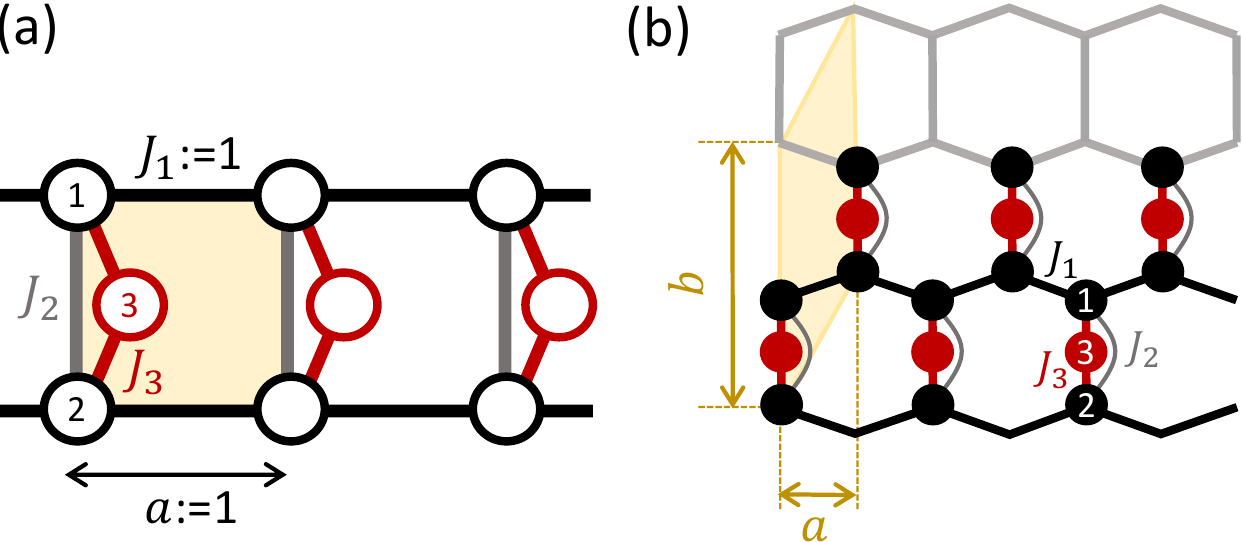}
% \vspace{-.2in}
\end{center}
\caption{The model system.
(a) Two-leg Heisenberg ladder with trimer rungs
    and couplings $J\equiv(J_1,J_2,J_3)$.
    A rung consists of three spin $S=\sfrac{1}{2}$ sites,
    where $m=1,2$ are the sites on the legs of the ladder.
    The presence of the additional {\it center} site on a rung
    (site $m=3$) coupled symmetrically to the leg
    sites via a finite coupling strength $J_3$
    induces frustration.
(b) The same model may be seen to represent a
    hexagonal brick lattice with three-site rungs, using $L_y=2$
    with periodic boundary condition in vertical direction.
    The lattice spacing of the 3-site unit cell
    (yellow shaded area) is taken as $a=1$ (horizontally,
    in either case), or $b$ [vertically, panel (b) only].
} \vspace{-.2in}
\label{fig:model}
\end{figure}

In this paper we study a model of a frustrated spin
$S=\sfrac{1}{2}$ Heisenberg ladder antiferromagnet
that is motivated by a quasi-one-dimensional (1D)
reduction of the trimer magnet Ba$_4$Ir$_3$O$_{10}$
\cite{Wilkens91,Stitzer02,Cao20}.
That material is a member of the hexagonal perovskite
family considered a potential host for quantum spin
liquid behavior \cite{Nguyen20}. It consists of a
layered structure with two-dimensional (2D) planes
where trimer units interconnected into a
quasi-hexagonal structure [cf. \Fig{fig:model}(b)].
The magnetism comes from trimer units that host three
Ir$^{4+}$ spin-half ions located within face-sharing
octahedra.
The dimensional reduction to 1D is partially
justified by experimental indications, and is
consistent with an extremely low Neel ordering
temperature $T_N = 0.2\, \mathrm{K}$ for the
material where the bandwidth of the spin
excitations by the Heisenberg couplings is of
several hundreds of Kelvin \cite{Cao20},
spanning nearly four orders of magnitude in energy
scales. This material may thus be instrumental
to the investigation of the long-standing
speculation that 2D frustrated magnets might
support quantum disordered states with neutral
spin-1/2 excitations known as
spinons~\cite{Balents_NP_07_frustration}.

Our interest in this system is driven by its unusual
spin arrangement as schematically depicted
in Fig. \ref{fig:model}. This
arrangement is conducive to several interesting
effects. In its classical Ising limit, the system
exhibits a frustration-driven ultra-narrow phase
crossover at finite temperature
\cite{Yin_MPT,Yin_icecreamcone}. For the quantum
case, as we will demonstrate in this work, the excitation
spectrum contains a soft gapless mode separated from
the other excitations by a gap in the limit of weak
interchain coupling $J_2,J_3 \ll J_1$. The main
contribution to the spectral weight of this mode comes
from the central spins on the rungs. These spins
nearly decouple from the system due to frustration,
and only experience an effective weak higher-order 
interaction amongst each other.
As far as the spins located on
the legs of the ladder are concerned, most of their
spectral weight is located at higher energies in
agreement with experimental observations. 

We point out that though the model looks like a
version of 3-leg ladder which is expected to be
equivalent to spin-1/2 chain and hence to be critical,
this equivalence does not hold throughout the entire
phase diagram. In the parameter range where the
interchain interaction is frustrated 
the ``orbital'' fluctuations are active and they
may lead to dimerization \cite{Majumdar69,White96,
Fouet06, Nishimoto09, Ivanov10}
which is absent in spin-1/2 chains.

The paper is organized as follows: In \Sec{sec:model}
we introduce the model. We then discuss first the strong
% (weak)
rung-coupling regime in \Sec{sec:strong-rung-cpl},
followed by the weak rung-coupling regime
in \Sec{seq:wk-rung-cpl}. Each % of these
contains an analytical treatment together with a
complimentary density matrix renormalization group
(DMRG \cite{White92,Schollwoeck11})
analysis including simulations of
the dynamical structure factor.
\Sec{sec:conclusions} contains conclusions.
\App{App:downfolding} gives additional background on the
downfolding to the effective low-energy Hamiltonian.
\Apps{app:meanfield} and \ref{app:absenceZ2}
discuss an alternative, even though not
physically realized, possibility of an intermediate
isotropic nematic phase with spontaneously broken rung
mirror symmetry due to frustration.

\section{Model}\label{sec:model}

We study the frustrated ladder model in \Fig{fig:model},
described by the Heisenberg Hamiltonian,
\begin{eqnarray} 
 H &=& \sum_i \Bigl[
   \sum_{m=1,2} \bigl(
     J_1\, \Ssite{i m}{\cdot}\Ssite{i+1,m}
  {+}J_3\, \Ssite{i m}{\cdot}\Ssite{i 3}
   \bigr)
 {+}J_2\, \Ssite{i 1} {\cdot}\Ssite{i 2}
 \Bigr]
\notag \\
 &\equiv& H^\mathrm{legs}(J_1) + 
 \sum_i H^\mathrm{rung}_i(J_2,J_3)
% \text{ ,}
\label{eq:model}
\end{eqnarray}
with isotropic spin interactions, where rung $i$
consists of three spin-half sites $m{=}1,2,3$
described by the spin operators $\Ssite{i,m}$. The
couplings $(J_1,J_2)$ describe the regular ladder of
two legs, whereas $J_3$ couples the two legs to a
third site $m=3$ present for each rung and referred to
as the {\it center site}, which thus frustrates the
direct coupling in between the legs. We consider
antiferromagnetic $J_i>0$, throughout. Moreover, we
assume $J_1=1$ as unit of energy, unless specified
otherwise, as well as $\hbar=1$.

The Hamiltonian \eqref{eq:model} has SU(2) spin
symmetry, as well as up-down mirror symmetry as shown
in \Fig{fig:model}(a), with the latter referred to as
rung inversion or \Ztwo parity symmetry. Within the
analytical or mean-field approach we assume periodic
boundary conditions (BCs), whereas in the DMRG
simulations we adopt open BCs, with the ladder
terminated by rungs $i=1$ and $N$. Overall, we are
interested in the thermodynamic limit $N\to\infty$. 

Taking $J_3$ to zero reduces the system to the
ordinary 2-leg ladder, except for the presence of
decoupled, and thus free center spins. The latter
would introduce macroscopic degeneracy. As will be
shown an approximate decoupling can also be achieved
by frustration that, in practice, may give rise to
spin freezing \cite{Zaliznyak99}. Also, one could
introduce a Heisenberg interaction $J_1'$ in between
nearest-neighbor center spins in \Fig{fig:model}(a),
thus resulting in a 3-leg ladder. But such a coupling
$J_1'$ is considered negligible here, except for the
discussion with \Fig{fig:3leg}. The motivation for
this is that the center spins may not necessarily be
nearest-neighbor (NN) spins, after all, as compared to
the leg spins. For example, one may assume that the
center spins in \Fig{fig:model}(a) have a 2-rung
periodicity, in that e.g., they point into and out of
the plane in an alternating fashion. Even more, when
viewed as a minimal model for the decorated brick
lattice in \Fig{fig:model}(b) assuming $L_y=2$, the
center sites are, indeed, very clearly far from being
NN sites.

\section{The limit of strong rung exchange}
\label{sec:strong-rung-cpl}

Let us start with focusing on a single rung consisting
of three sites as marked in \Fig{fig:model}. This is
relevant in the limit $J_1\to 0$, i.e., dominating
rung coupling, which reduces the system to (nearly)
decoupled trimers or triangles. Due to the SU(2) spin
symmetry, the state space of a rung can be reduced
from $d=2^3$ states to an effective dimension of
$d^\ast=3$ multiplets, having two $S=\sfrac{1}{2}$
multiplets and one multiplet with $S=\sfrac{3}{2}$.
The latter is already symmetric under rung inversion.
The two $S=\sfrac{1}{2}$ multiplets can be
symmetrized, where the first is symmetric and the
second asymmetric under rung inversion (by convention,
the $S=\sfrac{1}{2}$ rung multiplet space will be
always considered in this order). They will be
denoted by \Mtwo. The antisymmetric multiplet \Masym
forms a singlet across sites $m=(1,2)$, with a free
spin-half at site $m=3$. The symmetric multiplet
\Msym can be derived as a symmetric triplet on sites
$m=(1,2)$ that when fused with site $3$, also forms a
total rung spin $S=\sfrac{1}{2}$.
The eigenstates of a single rung thus are divided into
two groups: the low-energy space consisting of the two
spin-full ``orbitals'' \Mtwo at respective energies
$E^{(0)}_{\pm} = -\tfrac{\DeltaZ}{2} \pm \tfrac{1}{2}
(J_2-J_3)$, with $\DeltaZ \equiv \tfrac{1}{2} (J_2 +
2J_3)$, and the high-energy $S=\sfrac{3}{2}$ multiplet
at energy $+\tfrac{\DeltaZ}{2}$.

Frustration within each rung is therefore strongest
when $J_2 \approx J_3$. In this case, the two orbitals
\Mtwo become degenerate in energy. A finite detuning
$J_2 \neq J_3$ introduces an {\it orbital} splitting
by an energy exactly equal to $J_2-J_3$. This
motivates the dimensionless parameter,
\begin{equation}
   \alpha \equiv \tfrac{J_2 - J_3}{J_1}
\text{ .}\label{eq:alpha}
\end{equation}
The smaller the magnitude $|\alpha|$, the stronger the
frustration.

The excitation energy from the low-energy states to
the high-energy states is $\Delta=\DeltaZ \pm
\tfrac{\alpha J_1}{2}$. For low enough temperatures
$T$ satisfying $e^{-\Delta/T}\ll 1$ together with
$|\alpha|\ll 1$, the thermal population of the
high-energy $S=\sfrac{3}{2}$ multiplet vanishes.
With the high-energy states irrelevant to the low-energy
physics, we integrate it out by projecting the
Hamiltonian of two nearest-neighbor rungs including
their interaction along the legs into the low-energy
space formed by the multiplet space \Mtwo using the
many-body down-folding method \cite{White_NCT_02,
Yin_PRB_09_cuprates, Yin_downfolding_conf,
Yin_PRL_LaMnO3, Yin_PRL_Sr3CuIrO6, Yin_PRL_pyroxene}
based on Hubbard operators \cite{Hubbard_X_Operator}.
In order to make the physics more apparent, we find it
more convenient to represent these operators as
products of Pauli matrices acting in the spin and 
effective orbital sector (see \App{App:downfolding}
for details), as customary in theoretical studies of
manganites with colossal magnetoresistance and many
other materials with active orbital physics
\cite{Dagotto_PhysReports_01_manganites,
Moreo_Science_1999,Imada_RMP_98_MIT,
Yin_PRL_01_LaMnO3,Yin_PRB_07_EFE,
Yin_PRL_pyroxene,Yin_PRL_FeTe}.

Since the strengths of the bare projection of the
Hamiltonian for a nearest-neighbor pair of rungs and
the second-order perturbative terms are proportional
to $J_1$ and $J_1^2/\DeltaZ$, respectively
\cite{Yin_PRL_Sr3CuIrO6}, for strong rung-coupling
$J_1/\DeltaZ\ll 1$ it suffices to study the lowest
order.  This is described by the projected low-energy
Hamiltonian,
\begin{subequations}\label{eq:Heff}
\begin{eqnarray}
  \tfrac{1}{J_1}\, \Heff_\alpha
  = \tfrac{8}{9} \sum_i \left(
     \Sspin_i\cdot\Sspin_{i+1}
  \right) \otimes \mathbb{T}_{i,i+1}
    + \alpha  \sum_{i=1}^{N}
     \Torb_i^z \qquad
\label{eq:Heff:1}
\end{eqnarray}
where given the 2-leg ladder with center spins
on the rungs we write
$\mathbb{T}_{i,i+1} = \mathbb{T}_{i,i+1}^{(2)}$,
we have
\begin{eqnarray}
  \mathbb{T}_{i,i+1}^{(2)} \equiv \tfrac{1}{4}
   + \tfrac{1}{2} (\Torb_i^z {+} \Torb_{i+1}^z)
   + \Torb_i^z\Torb_{i+1}^z + 
     3\Torb_i^x \Torb_{i+1}^x
\text{ ,}\quad\label{eq:Heff:2}
\end{eqnarray}
\end{subequations}
Here $\Sspin_i^a \equiv \tfrac{1}{2} \sigma_i^a$ and
$\Torb_i^a \equiv \tfrac{1}{2}\tau_i^a$ are effective
spin and orbital spin-half operators, respectively,
with $\sigma^a$ and $\tau^a$ Pauli matrices with
$a\in\{x,y,z\}$. These form the direct product space
$\sigma \otimes \tau$ that acts on rung $i$. Matter
of fact, the new spin operators exactly correspond to
the total spin operator on a rung, $\Sspin_i \equiv
\Ssite{i}^{\mathrm{tot}} {\equiv} \sum_{m=1}^3
\Ssite{i m}$, which, once projected onto the
low-energy spin sector, indeed, represent a plain
proper spin operator acting on a $S=\sfrac{1}{2}$ spin
degree of freedom.

The last term in \Eq{eq:Heff:1} is nothing but the
aforementioned ``orbital'' splitting of $\alpha J_1$.
It now functions as an effective magnetic field
applied on the $\tau$ pseudo-spins along the
$z$-direction. It is offset by the linear $\Torb^z_i$
term in \Eq{eq:Heff:2}. The prefactor can be roughly
estimated via a mean-field value for a decoupled
Heisenberg chain \cite{Hulthen_Heisenberg_GS,
Bethe_ansatz_1931}, resulting in [cf. \App{app:meanfield}]
\begin{eqnarray}
  \alpha_0 \equiv - \tfrac{8}{9}
  \langle \mathbf{\Sspin}_i\cdot\mathbf{\Sspin}_{i+1}\rangle
  \approx  \tfrac{8}{9} (\ln{2} - \tfrac{1}{4}) = 0.394
\text{ .}\label{eq:alpha0}
\end{eqnarray}
Therefore only if $\alpha\approx\alpha_0$, the
effective magnetic field becomes zero in the orbital
sector. This offset also approximately agrees with the
full many-body calculation, where the DMRG simulation
in \Fig{fig:dim_gap}(c) shows that $\alpha_0$
renormalizes to a slightly smaller value of $0.341$.

The orbital magnetization $\langle \Torb^z_i \rangle$
or $\langle \Torb^x_i \rangle$ can be directly related
to the intra-rung spin-spin correlations,
\begin{subequations} \label{SdotS:corr}
\begin{eqnarray}
   C_{1 2}^{(i)}
   &\equiv&
     \langle \Ssite{i 1}\cdot \Ssite{i 2} \rangle \hspace{.46in}
   =-\tfrac{1}{4} + \langle \Torb^z_i \rangle
\label{eq:C12} \\
   C_{3+}^{(i)}
   &\equiv&
     \langle (\Ssite{i 1}
            + \Ssite{i 2} ) \cdot \Ssite{i 3} \rangle
   =-\tfrac{1}{2} - \langle \Torb^z_i \rangle
\label{eq:C13} \\
   C_{3-}^{(i)}
   &\equiv&
     \langle (\Ssite{i 1}
            - \Ssite{i 2} ) \cdot \Ssite{i 3} \rangle
   = \quad\,\sqrt{3}\, \langle \Torb^x_i \rangle % \text{ ,}
\label{eq:Tx}
\end{eqnarray}
\end{subequations}
where $\langle ...\rangle$ denotes thermodynamic
average. \EQ{eq:C12} shows that $\langle \Torb^z_i
\rangle$ measures whether the two leg spins (1,2) are
ferromagnetically or antiferromagnetically correlated.
Matter of fact, $\langle \tau^z_i \rangle \equiv
2\langle \Torb^z_i \rangle$ measures the rung parity
\Ztwo, where based on \Eq{eq:C12}, $\tau^z_i$ acts
like a swap operator for the two leg sites.
Conversely, $\langle \Torb^x_i \rangle$ measures the
\Ztwo symmetry breaking between the leg spins if
present. As seen from \Eq{eq:Tx}, a non-zero value
indicates a spontaneous breaking of the mirror
symmetry between the upper and lower leg.

As an aside we note that when the site-specific spin
operators themselves are fully projected into the
low-energy space, caveats apply, e.g., for sum rules.
Since $\Sspin _i = \mathbf{S}^\mathrm{tot}_\mathrm{i}$
is fully constrained to the $S=1/2$ spin sector, one
obtains $\Sspin _i^2 = \tfrac{3}{4}$. However, if the
site-specific spin operators $\Ssite{i m}$ themselves
are fully projected to the low-energy
$S{=}\sfrac{1}{2}$ space, then $\sum_{m=1}^3 \Ssite{i
m}^2 = \tfrac{5}{4}$ (and not $3\cdot \tfrac{3}{4}$,
as this misses weight not of interest from
intermediate excitations into the high-energy
$S{=}\sfrac{3}{2}$ multiplet), such that the sum
rule becomes $\bigl(
  \Ssite{1}\cdot \Ssite{2} +
  \Ssite{1}\cdot \Ssite{3} +
  \Ssite{2}\cdot \Ssite{3} \bigr)_i
= \tfrac{1}{2}\, ( \tfrac{3}{4}{-}\tfrac{5}{4} ) 
=-\tfrac{1}{4}$.
In the absence of intermediate truncation in the spin
operator products as with \Eqs{SdotS:corr} above, this
reads $C_{12} + C_{3+} = -\tfrac{3}{4}$, instead.

The effective Hamiltonian \eqref{eq:Heff} only
includes nearest-neighbor terms derived from bare
projection which, at first glance, may be taken as
indication for a uniform ground state. In addition,
one may also includes next-nearest neighbor (NNN)
interactions via second order perturbation. This
translates the local rung frustration of the original
ladder into frustration along the chain in the
effective model. Such NNN interactions, while they
leave the effective Hamiltonian translationally
invariant, can be expected to generate dimerization as
a relevant perturbation. This can give rise to
spontaneous breaking of the translational symmetry
along the chain \cite{Majumdar69,White96}. Based on
second order perturbation, such a symmetry breaking,
however, should diminish in the limit of strong rung
couplings $J_2,J_3 \gg 1$.

Nevertheless, as will be seen in the DMRG analysis
below, the lowest-order projected Hamiltonian in
\Eq{eq:Heff} itself already gives rise to
dimerization. Being at lowest order, the resulting
dimerization also does not diminish but remains
sizeable in the limit of strong rung couplings
$J_2,J_3 \gg 1$. This suggests that spin and orbital
degrees of freedom remain intrinsically entangled, and
cannot be mean-field decoupled. The frustration of
the spins on each rung in the original model is
present via the (near) degeneracy of the two
multiplets \Mtwo. One may argue that the decoupled
spin chains described by the first term only in
\Eq{eq:Heff:2} are subjected to relevant effective NNN
order terms based on the remainder of the interactions
in \Eq{eq:Heff:2}. Therefore, overall, frustration is
already intrinsic also to the effective projected
Hamiltonian \eqref{eq:Heff:2}.

Interestingly, dimerization as found in our DMRG
simulations has been reported on an isotropic 3-leg
Heisenberg ladder in \cite{Nishimoto09}. Translated
to our model, this would turn on the coupling also for
nearest-neighbor center sites ($m=3$). Taking it
equally strong as for the initial two legs having
$J_1$, then following the same down-folding procedure
above, one obtains instead of \Eq{eq:Heff:2} the
modified effective Hamiltonian in the orbital sector,
\begin{eqnarray}
  \mathbb{T}_{i,i+1}^{(3)} \equiv \tfrac{3}{8}
   + 3 \bigl( \Torb_i^x \Torb_{i+1}^x
   +   \Torb_i^z \Torb_{i+1}^z \bigr)
\text{ .}\quad\label{eq:Heff:3}
\end{eqnarray}
As compared to the 2-leg case in \Eq{eq:Heff:2}, the
linear terms in $\Torb^z$ disappeared [hence one also
expects no offset here to the orbital magnetic field
as estimated in \Eq{eq:alpha0}]. Also the $\Torb^z
\Torb^z$ term got strengthened, making it equally
strong as the $\Torb^x \Torb^x$ term which kept its
prefactor unchanged. If one were to analyze the
orbital sector effectively decoupled from the spin
sector, this would result in plain Fermionic tight
binding chain after Jordan-Wigner transformation. On
the contrary, however, also the 3-leg ladder above
features dimerization, instead \cite{Nishimoto09}.
This emphasizes the strongly correlated interplay
between spin and orbital degrees of freedom. We will
show below by continuously turning on the NN
Heisenberg coupling on the center spins [cf.
\Fig{fig:3leg}] that the spin gap observed with
dimerization in the system never closes on the way
making an isotropic 3-leg ladder with the same
coupling $J_1$ on all three legs. This suggests, that
the underlying physics is identical.

\subsection{Preliminary discussion}

We proceed to discuss the physics of the effective
strong coupling Hamiltonian. The symmetries which can
be spontaneously broken in the ground state are the
\Ztwo symmetry between the chains and the
translational, or to be more precise, the inversion
symmetry along the chains. Qualitative considerations
suggest a possibility of the following $T=0$ phases.
First, there are two diagonal, ``orbital''-ordered
phases with $\langle\Torb^z\rangle>0$ and
$\langle\Torb^z\rangle<0$, respectively, which can
coexist with translational symmetry breaking. There
is the possibility of a nematic phase with
spontaneously broken \Ztwo symmetry, having
$\langle\Torb^x\rangle\neq 0$. It is nematic, since
with \Eq{eq:Tx} the local order parameter would
consist of \Ztwo symmetry-breaking variations in the
energy density described by scalar products of spins
with the SU(2) spin symmetry itself preserved. The
nematic order may coexist with translational symmetry
breaking.

The diagonal phases appear at strong effective field
$|\alpha| \gg 1$ [cf. \Eq{eq:alpha}], while also $J_2,
J_3 \gg 1$ $(=J_1)$. Then quantum orbital
fluctuations in the $xy$ orbital plane are suppressed.
For $\alpha\gg 1$, i.e., dominant $J_2 \gg J_3$, the
effective field via the last term in \Eq{eq:Heff:1}
aligns $\langle \Torb_i^z\rangle {\simeq}
-\frac{1}{2}$. By \Eq{eq:C12}, this results in the
strongest possible antiferromagnetic correlation for
the leg spins (1,2), such that they form an
approximate singlet ($S=0$), while the center spin
becomes nearly decoupled. Overall, this is precisely
the antisymmetric rung multiplet \Masym.
On the other hand, for $\alpha \ll -1$, i.e.,
dominating coupling to the center spin, $J_3 \gg J_2$,
the effective field in \Eq{eq:Heff:1} aligns $ \langle
\Torb_i^z\rangle \simeq \frac{1}{2}$. Again by
\Eq{SdotS:corr}, this shows that here the leg spins
align ferromagnetically such that they form an
approximate triplet ($S=1$) with antiferromagnetically
aligned center spin. This is nothing but the symmetric
rung multiplet \Msym.

In the latter diagonal phase for $\alpha\ll-1$, the
spin dynamics in \Heff is described by a simple
single-chain spin-half Heisenberg model in terms of
the symmetric multiplet \Msym and coupling strength of
order $J_1$. As will be shown below, also the first
diagonal phase ($\alpha\gg 1$) reduces to an effective
spin-half Heisenberg model in terms of the
antisymmetric multiplet \Masym. There, however, this
translates into a Heisenberg chain of weakly coupled
center spins, such that in this case the coupling
strength, and with it the energy scale of the spin
dynamics, becomes vanishingly small for $\alpha \gg
1$.

The above analysis indicates that there may exist a
quantum critical point (QCP) in the regime of weak
$|\alpha|<1$ (i.e., strong spin frustration) that
separates the two phases with antiferromagnetic and
ferromagnetic correlations, respectively.
Alternatively, there is also the possibility of a
nematic phase for small $\alpha$ whose phase
boundaries would require two QCPs where the nematic
order vanishes. The latter is suggested by a
semi-meanfield approach as discussed in
\App{app:meanfield}. However, based on the detailed
DMRG analysis presented below, neither turns out to
capture the low-energy regime. Instead, the system
favors a spontaneously broken translational symmetry
with dimerization along the ladder that smoothly
connects the regime $\alpha \ll -1$ to $\alpha\gg 1$,
as will be demonstrated next.

\subsection{Dimerization}

\begin{figure}[tb!]
\begin{center}
\includegraphics[width=1\linewidth]{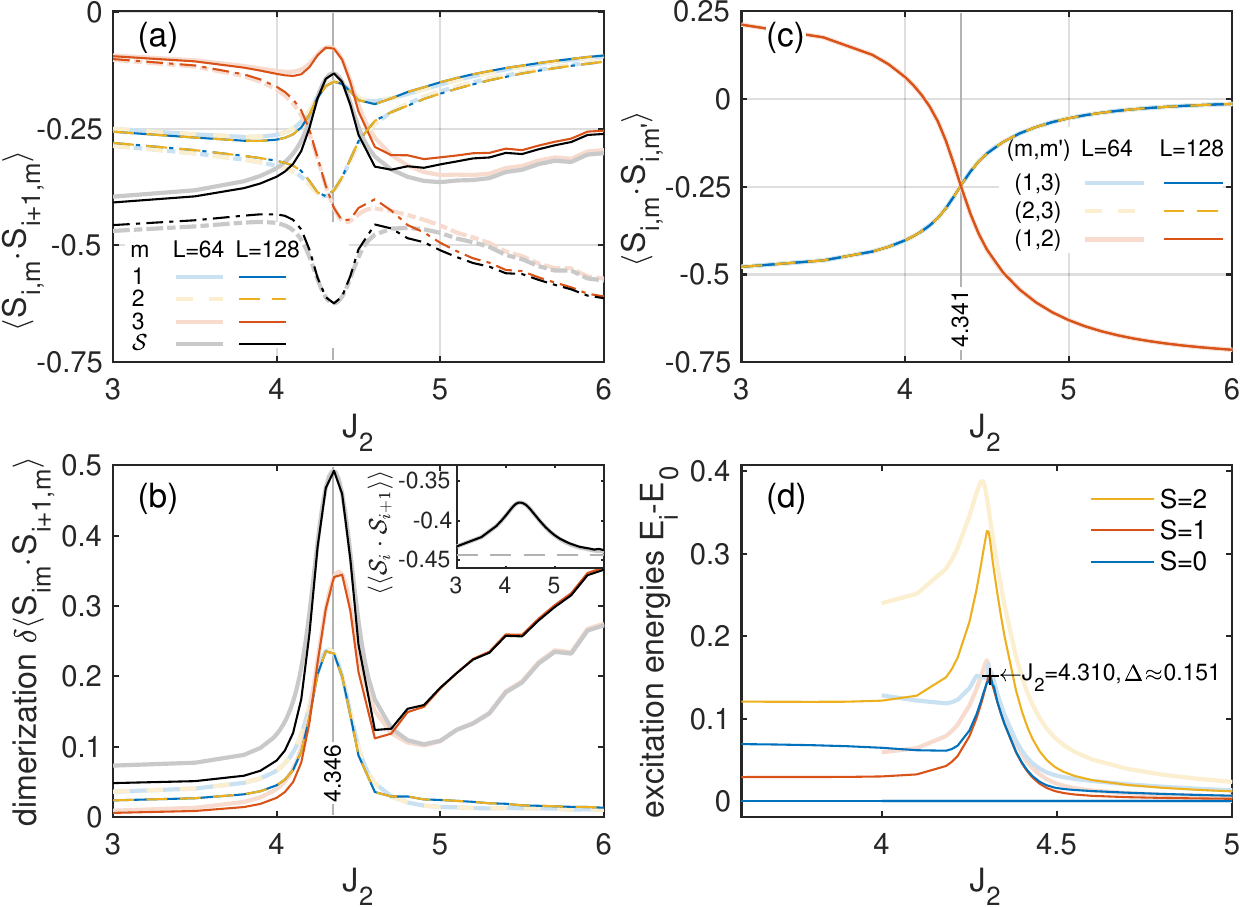}
\end{center}
\caption{
   Dimerization in the ladder model of Hamiltonian
   \eqref{eq:model} in the strong rung-coupling limit,
   having $J{=}(1,J_2,4)$ vs. $J_2$
   based on DMRG ground state simulations
   for arbitrary but fixed $J_2$ for
   $L=64$ (light thick lines) $L=128$ (color matched
   thin dark lines).
   (a) NN interaction energies along the chain
   in the system center, showing even (odd) bonds
   around the system center separately as individual
   curves in solid (dashed-dotted), respectively.
   The data includes NN interaction in terms of the original
   spins w.r.t. to sites $m=1,2,3$, but also 
   of $\Sspin_i \equiv {\bf S}_i^\mathrm{tot}$,
   as indicated with the legend.
   (b) Same as (a), also sharing the same legend,
   but plotting the difference 
   between the even and odd bonds along the chain.
      The inset shows the average over even and odd 
   bonds, denoted by $\langle\langle .. \rangle\rangle$,
   vs. $J_2$ for the $\langle\Sspin_i\cdot \Sspin_{i+1}\rangle$
   data in (a).
   Here again the color matched black line
   refers to $L=128$, whereas
   the lighter gray line (mostly underneath the black
   line) refers to $L=64$.
   The horizontal guide at the bottom of the inset
   indicates the analytically known expectation value
   for a plain spin-half Heisenberg chain,
   $\tfrac{1}{4} - \ln(2)$ [cf. \Eq{eq:alpha0}].
   (c) NN interaction energies within a rung in
   the system center. Same analysis as in (a),
   but here the data from even / odd rungs
   lies indistinguishably on top of each other.
   (d) Targeting lowest-energy states in global
   SU(2) spin sectors as indicated in the legend.
} \label{fig:dim_gap}
\end{figure}

\begin{figure}[tb!]
\begin{center}
\includegraphics[width=1\linewidth]{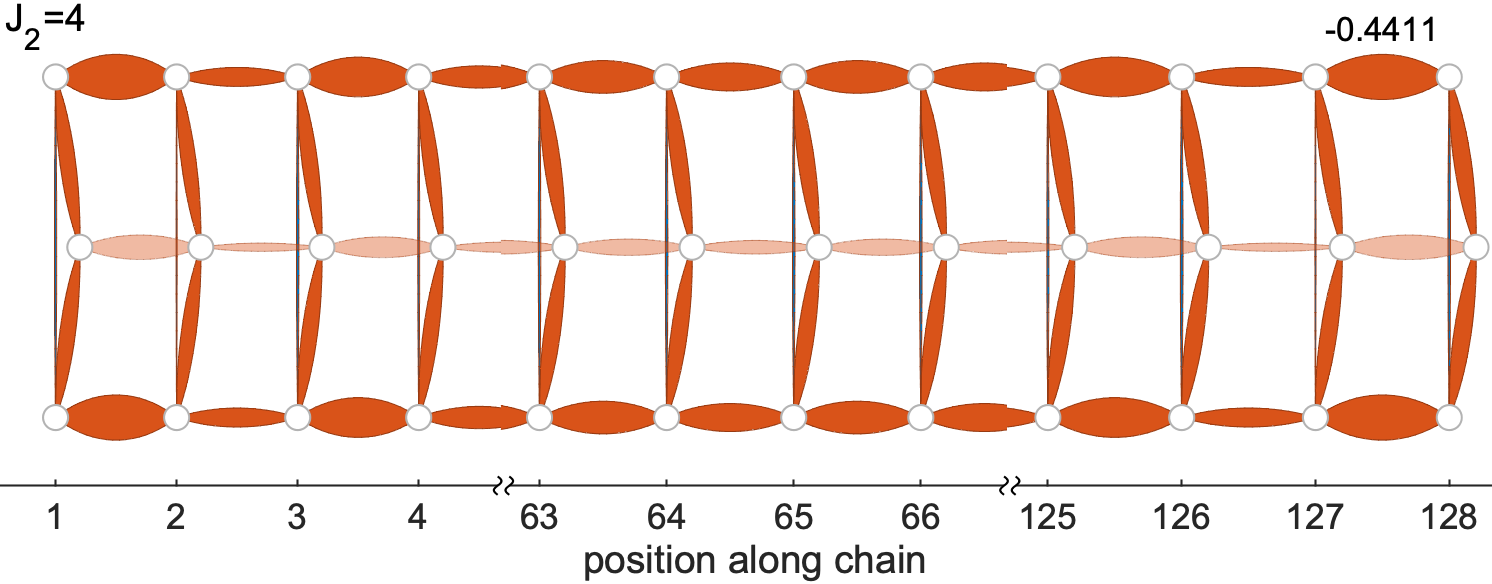}
\includegraphics[width=1\linewidth]{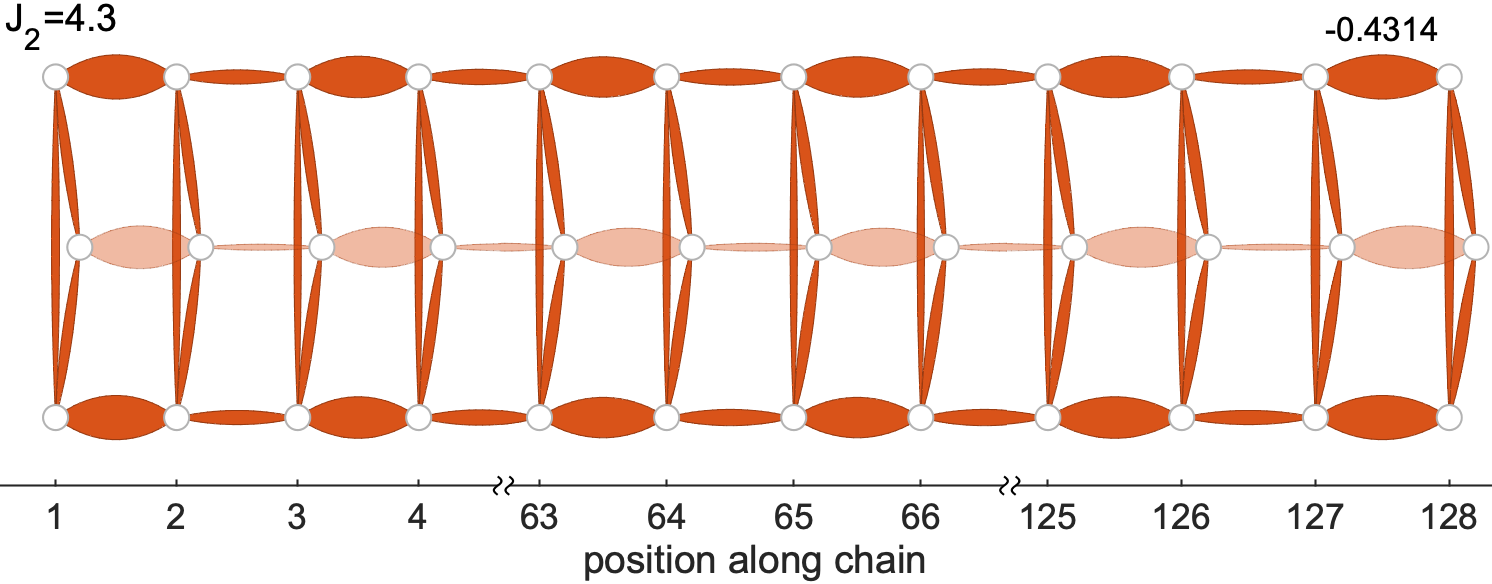}
\includegraphics[width=1\linewidth]{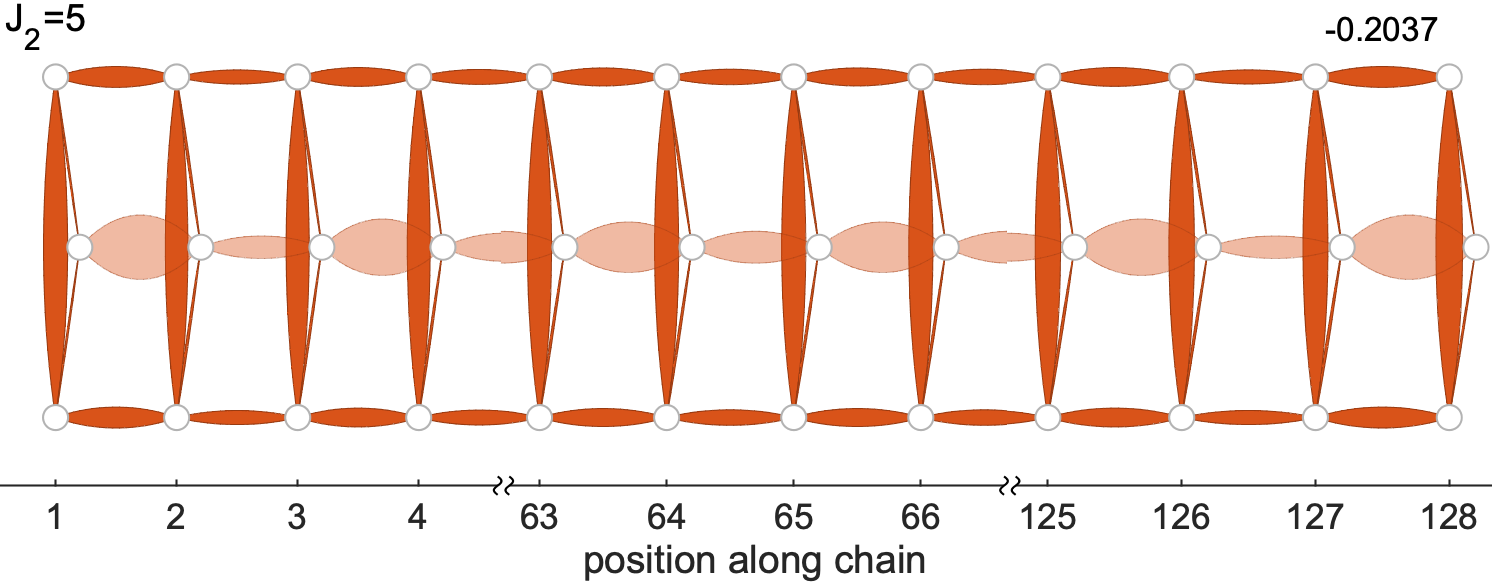}
\end{center}
\caption{
   Snapshots of NN bond strength $C_{m m'}^{(i,i')}
   \equiv \langle {\bf S}_{i m} {\cdot} {\bf
   S}_{i',m'} \rangle$ in the ladder model of
   Hamiltonian \eqref{eq:model} for $J{=}(1,J_2,4)$
   [same as in \Fig{fig:dim_gap}] with $J_2$ as
   specified with the panel. The NN bond strength is
   drawn to scale proportional to the bond width (see
   value for the bond at the upper right of each panel
   for reference). The NN interactions between center
   spins is shown semitransparent (light colors) to
   indicate that no interactions are present in the
   Hamiltonian for these bonds. All bonds are of the
   same color, and hence of the same negative sign,
   thus being antiferromagnetically correlated. The
   data is for an $L=128$ ladder, showing left
   boundary, center, and right boundary, with the
   intermediate ranges cropped as indicated with each
   horizontal axis.
} \label{fig:snapshots}
\end{figure}

In this section we present extensive DMRG
\cite{White92,Schollwoeck11} ground state simulations
on the two-leg ladder model in \Eq{eq:model}, as well
as in its projected version in \Eq{eq:Heff}. The
results are overall consistent, e.g., in that the
total weight in the reduced density matrix for $J_3=4$
in the local $S=\sfrac{3}{2}$ rung multiplet remained
below $0.01$, throughout. Here we use uniform ladders
with open boundary conditions for $J=(1,J_2,4)$ where
we scan $J_2$ and subsequently combine the data from
the system center for each DMRG run at fixed $J_2$.
Our results with focus on dimerization are summarized
in \Fig{fig:dim_gap}. Snapshots of the NN
spin-interactions are shown in \Fig{fig:snapshots} for
$J_2=4$, $4.3$, and $5$. The DMRG data for these
snapshots was obtained for a system size of $L=128$
rungs, with very minor variations as compared to
$L=64$, as seen in \Fig{fig:dim_gap}(a-c). For
clarity, we only show left end, center, and right end
of the ladder, with the intermediate regions cropped
as indicated with the lower axis sets. This
demonstrates that the dimerization is well-established
and uniform along the entire system.

Figures\,\ref{fig:dim_gap}(a,b) analyze the NN spin
correlations along the ladder, $C_{m m'}^{(i,i+1)}
\equiv \langle {\bf S}_{i m} {\cdot} {\bf S}_{i+1,m'}
\rangle$, whereas \FIG{fig:dim_gap}(c) shows the
perpendicular ones, i.e., within rungs. These
interactions are computed based on the actual sites
($m{=}1,2,3$), but in \Fig{fig:dim_gap}(a,c) also in
terms of the effective spin operator $\Sspin_i$ (black
line). By plotting data separately for even from odd
bonds in the system center of the ladders analyzed,
dimerization is absent if these curves lie on top of
each other [e.g., \Fig{fig:dim_gap}(c)]. Dimerization
develops where the curves split as in
\Fig{fig:dim_gap}(a), where \Fig{fig:dim_gap}(b) plots
the actual difference. Therefore for given parameter
setting, dimerization starts around $J_2 \gtrsim
J_3=4$ [cf. \Fig{fig:dim_gap}(a,b)]. It develops a
pronounced maximum around $J_2 \sim 4.3$
[\Fig{fig:dim_gap}(b)] and drops again thereafter up
to $J_2\sim 4.6$.
The dimerization `bubble' that opens between even and
odd bonds in \Fig{fig:dim_gap}(a) is absent for
\Fig{fig:dim_gap}(c) which analyzes the three bonds
{\it within} a rung. The latter data is uniform when
going from one rung to the next. Therefore
dimerization, and correspondingly spontaneous symmetry
breaking, only exist along the legs, but not within
the rungs.

The dimerization observed in the spin-spin
correlations for $J_2 \in [4,4.5] \gtrsim J_3$, and
therefore $\alpha \in [0,0.5]\gtrsim 0$ goes hand in
hand with the appearance of a small but
well-established spin-gap $\DeltaS \lesssim 0.15$ as
shown in \Fig{fig:dim_gap}(d). There by simultaneously
targeting multiple lowest-energy states in various
global SU(2) symmetry sectors, we find that both, the
singlet- and triplet gap are maximal for $J_2 \simeq
4.31$ and already well-converged to the
aforementioned value for $L=64$ (light colors) as
compared to $L=128$ (strong colors). While the ground
state evolves smoothly, the excited states feature a
sharp kink which suggests a crossing of state spaces.
This is natural bearing in mind that the many-body
Hilbert space can be partitioned into states that are
either symmetric or antisymmetric under rung inversion
symmetry and where their presence in the low-energy
regime is expected to be reversed for $J_2$
significantly larger or smaller as compared $J_3$.

The location of the maximal spin gap well coincides
with the crossing of weights for the symmetric and
antisymmetric rung-multiplet, as reflected by the
crossing of the lines in \Fig{fig:dim_gap}(c): that
crossing exactly occurs at the point where the
symmetric and antisymmetric rung multiplet, \Msym and
\Masym, gain equal weight, since with \Eq{SdotS:corr}
for $m\neq m'$, $\langle {\bf S}_{i m} \cdot {\bf
S}_{i,m'} \rangle = -0.25$ when $ \langle \Torb^z_i
\rangle = 0$. As indicated with
\Fig{fig:dim_gap}(c), the crossing occurs at $J_2
\simeq 4.341$, i.e., $\alpha_0^\mathrm{DMRG} \simeq
0.341$ which thus slightly reduces the mean-field
estimate for $\alpha_0$ in \Eq{eq:alpha0}.
For $J_2\lesssim 4.2$ the gap diminishes and dissolves
within strong finite-size effects. The system appears
critical and non-symmetry broken for $J_2\lesssim
J_3=4$ [e.g. see center region in upper panel of
\Fig{fig:snapshots}], even though based on the DMRG
data we cannot exclude that a small but finite gap
persists even for $1 \ll (J_2 < J_3)$.

The situation for large $J_2 \gtrsim 4.6$ differs as
compared to the case of small $J_2<J_3=4$. By looking
at \Fig{fig:dim_gap}, one notices two points: (i) the
finite-size spacing in \Fig{fig:dim_gap}(d) is much
smaller for large $J_2\gtrsim 4.6$ as compared to
$J_2<4$, and (ii) while the dimerization in
\Fig{fig:dim_gap}(b) diminishes on the actual legs of
the ladder ($m=1,2$), the dimerization starts to grow
again for the center spins for $J_2 \gtrsim 4.6$ (see
also lower snapshot in \Fig{fig:snapshots} for
$J_2=5$).
Point (i) is fully consistent with the earlier
discussion that for $\alpha \ll -1$ ($\alpha\gg 1$),
which in the present case roughly corresponds to
$J_2\lesssim J_3=4$ ($J_2 \gtrsim 4.6)$, respectively,
the symmetric rung multiplet \Msym (or antisymmetric
\Masym) dominate the rung state space. This is
clearly visible in the upper as compared to the lower
snapshot in \Fig{fig:snapshots}: the upper snapshots
ties in all three spins on a rung based on
antiferromagnetic correlations routed through the
center spin. In contrast, the lower snapshot directly
couples the leg spins hence resulting in a dominant
\Masym, which eventually results in these orbitals
being gapped out, akin to a rung singlet phase in the
plain Heisenberg ladder \cite{Wb18_SUN}. The residual
center spins, however, only experience a very weak
indirect coupling amongst each other via higher-order
perturbative processes. Their effective spin-spin
interaction diminishes to zero for $J_2\gg J_3$, in
qualitative agreement with the finite-size level
spacing see in \Fig{fig:dim_gap}(d).

Point (ii) is {\it a-priori} unexpected. While all our
DMRG data is very well-converged to start with, e.g.,
even also for all the $L=128$ data the ground state
energy is converged to well below $10^{-6}$ relative
accuracy, throughout, there is room to believe that
the eventual increase of the dimerization with the
center spins in \Fig{fig:dim_gap}(b) is a numerical
artifact.
Matter of fact, the DMRG simulations for $J_2\gtrsim
4.6$ were difficult to start with in that random
initialization also randomizes the (very) weakly
coupled center spins. This becomes very difficult to
get rid of towards a more uniform ladder later, in
that DMRG may be stuck within certain initial
antiferromagnetic spin clusters with domain walls in
between. Hence for $J_2\gtrsim 4.6$, the DMRG was
(also) initialized with a drastically down-sampled
ground state obtained for smaller $J_2\sim 4.3$. For
the larger $J_2$ values where a randomized starting
state could still be afforded, the resulting data was
overall consistent. Nevertheless, as seen from
\Fig{fig:dim_gap}(b), the $L=64$ shows a
systematically smaller dimerization for $J_2\gtrsim 5$
which may be attributed to the fact the $L=64$ data is
still overall systematically somewhat better converged
than $L=128$. So one may take this as a first
indication that the dimerization seen with the center
spins for large $J_2$ shows a tendency to become
smaller or even diminish altogether. Besides, the
data for large $J_2$ also shows some minor irregular,
noisy behavior vs. $J_2$ for either system length $L$.
This is mainly also attributed to the quick decoupling
of the center spins with increasing $J_2$. For similar
reasons, the dimerization on the center spins may also
be strongly influenced still by the presence of the
open boundaries. The precise fate of the dimerization
for large $J_2$ therefore remains open, but there is
room to believe that it diminishes for large $J_2$
also for the center spins eventually. In this sense,
in what follows we only refer to the intermediate
range $J_2 \sim J_3 + [0,0.5]$ with $J_3 \gg J_1=1$
as the (clearly) dimerized regime with the precise
boundaries of this phase left for future studies.

The averaged correlations $\langle\langle \Sspin_i
\cdot \Sspin_{i+1} \rangle\rangle$ including both,
even and odd bonds is shown in the inset of
\Fig{fig:dim_gap}(b). For $J_2$ far detuned from
$J_3=4$, this approaches the analytical value known
for the plain spin-half Heisenberg chain indicated by
the horizontal line. This clearly supports the
overall picture that in the strong rung-coupling
limit, the system effectively reduces to a plain
spin-half Heisenberg chain, either in the symmetric
or antisymmetric rung multiplet, \Msym or \Masym, for
$J_2 \lesssim J_3$ or $J_2 \gtrsim J_3+0.6$,
respectively.

\begin{figure}[tb!]
\begin{center}
\includegraphics[width=0.9\linewidth]{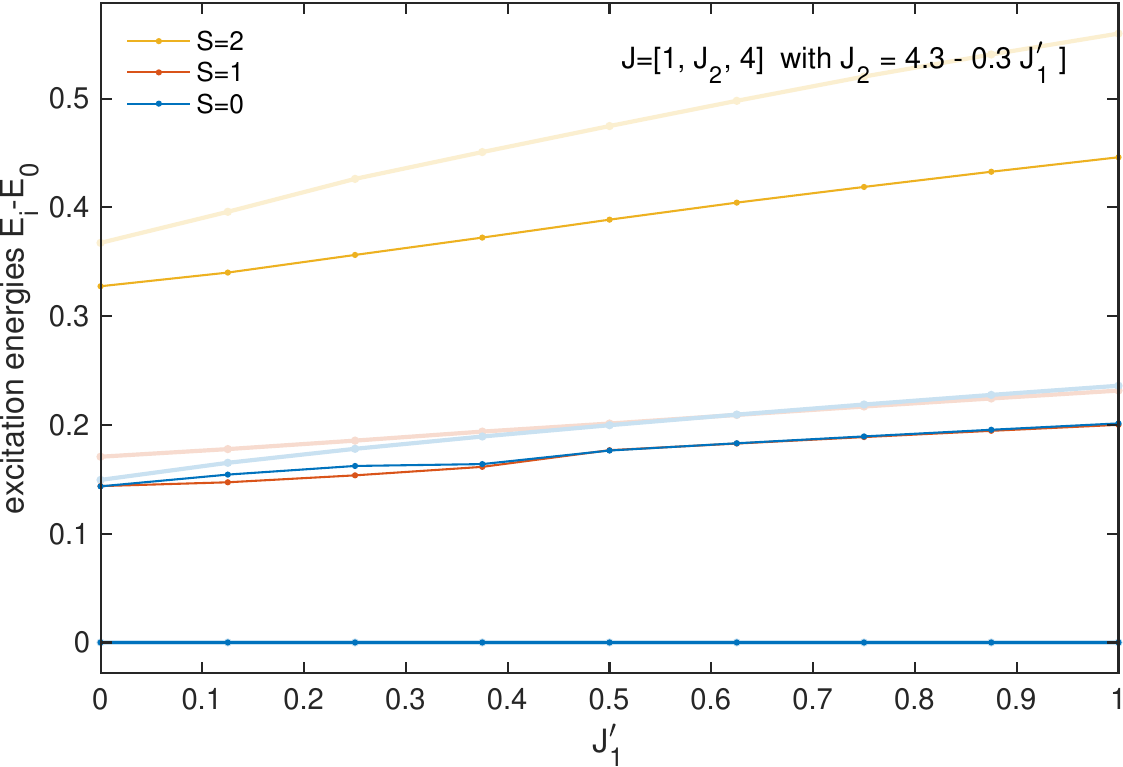}
\end{center}
\caption{
   Lowest-energy eigenstates of model \eqref{eq:model}
   in the strong rung-coupling regime except that
   here, in addition, also a NN Heisenberg coupling
   $J'_1 \in [0,1]$ (horizontal axis) between the
   center spins was turned on. Therefore $J'_1=0$
   well corresponds to the projected 2-leg ladder in
   \Eq{eq:Heff:2}. $J_2$ was tuned with $J'_1$ as
   indicated in the panel, such that $J'_1=1$
   corresponds to the uniform 3-leg model with the
   projected low-energy orbital Hamiltonian as in
   \Eq{eq:Heff:3}. While the full, i.e.,
   non-projected rung state space was present in the
   simulations, the total weight of the
   $S=\sfrac{3}{2}$ multiplet was $\lesssim0.01$,
   throughout. Light colors are for $L=64$, whereas
   darker colors are for $L=128$, similar to
   \Fig{fig:dim_gap}(d). States are color-coded
   according to their global SU(2) spin sectors as
   indicated in the legend.
} \label{fig:3leg}
\end{figure}

\subsection{Connection to dimerized regime in 3-leg tube}

The model of interest in this work is the 2-leg ladder
in \Eq{eq:model}, or its projected version in
\Eq{eq:Heff}. Nevertheless, in the dimerized regime
as in the center snapshot in \Fig{fig:snapshots}, the
center spins virtually correlate the same way as the
leg spins, despite that there is no direct coupling in
between the center spins in the Hamiltonian at all
[hence these bonds were depicted in semitransparent
(light) colors]. Based on this, one may suspect that
the dimerized phase persists even if a NN coupling is
explicitly turned on also in between the center spins
in the Hamiltonian. The resulting projected model
results in the altered orbital Hamiltonian
\eqref{eq:Heff:3}.

\FIG{fig:3leg} tracks the spin gap while turning on an
explicit NN coupling $J'_1$ in between center spins,
starting around the maximal spin gap in the 2-leg
model at $J = [1,4.3,4]$ at $J'_1=0$ [cf.
\Fig{fig:dim_gap}]. Because the 3-leg Hamiltonian in
\eqref{eq:Heff:3} has no linear offset to the orbital
magnetic field as discussed with \Eq{eq:alpha0}, at
the same time as turning on $J'_1$, $J_2$ is tuned
linearly towards $J_3$, i.e., $\alpha=0$, as indicated
with \Fig{fig:3leg}. With this, $J'_1=1=J_1$
corresponds to 3-leg `tube' 
\cite{Fouet06,Nishimoto09,Ivanov10}
with three equivalent
legs, having $J=[1, 4, 4]$. As evident from
\Fig{fig:3leg}, the spin gap never closes, it even
gets enhanced as $J'_1$ is turned on. Hence the
dimerization of the 3-leg tube observed in
Ref.\,\onlinecite{Nishimoto09} has the same physical
character as the dimerized phase observed for the
2-leg model here. Ref. \citenum{Nishimoto09} analyzed
the 3-leg tube for any $J_2=J_3$ relative to $J_1$,
which in the present case translates to
$\alpha^\mathrm{eff}\equiv \alpha-\alpha_0 = 0$. They
argued that this model is always gapped and dimerized
due to spin-frustration. Therefore \Fig{fig:3leg}
shows that the dimerized regime seen in our model has
the same physical origin, namely a frustration induced
spin-Peierls transition \cite{Kawano97}. As we will
demonstrate below, the spin-Peierls character of the
dimerized phase is supported by the analytic
calculations.

\subsection{Dynamical properties and crossover of spinon continua}
\label{sec:DSF}

The dynamical structure factor (DSF) examines the
energetics of spin-spin correlations. Here we use it
in the form,
\begin{eqnarray}
   S_{m m'}(k,\omega)
 = \sum_i e^{-\iu k x_i} \!\int \! dt\,
   e^{\iu \omega t} \, S_{m m'}(x_i,t)
\text{ ,}\label{eq:Smm'}
\end{eqnarray}
where we only consider momentum $k$ along the ladder,
yet site-specific, and hence with real-space resolution
along the `vertical' direction within a rung.
Here $x_i$ refers to the horizontal distance along the
ladder using unit lattice spacing, $x_i = i$,
with $m$ and $m'$ the local site indices within
a rung, having $S_{m m'}(x_i,t) \equiv
\bigl\langle \Ssite{i m}(t) \cdot \Ssite{0 m'}(0)
\bigr\rangle$ with site spins $\Ssite{i m}$
as in \Eq{eq:model} with SU(2) spin symmetry intact.
Here $S_{0 m'}$ refers to site $m'$ on
a reference rung at location $i'=0$. In the present
DMRG context, using open BCs, this always refers
to a site on the center rung of the system.
The DMRG prescription is then as follows: 
one performs real-time evolution \cite{White04,Daley04},
followed by double-Fourier transforms. To be specific,
we subtract a static long-time background,
perform zero padding in real space, followed by
Fourier transform to momentum space. After careful
linear prediction \cite{Barthel09} of $S(k,t)$ in time,
the system is then also Fourier transformed to
frequency space, followed by a final weak broadening
to remove artificial speckles from pushing linear
prediction. We emphasize that linear prediction
in momentum space, and thus mixed coordinates $(k,t)$
is crucial, since for fixed $k$ significantly fewer
frequencies occur within $S(k,t)$. This is in stark
contrast to $S(x,t)$ which has all frequencies
from the entire DSF spectrum present which then
results in delayed, light-cone like dynamics that is
ill-suited for linear prediction.
 
The DSF obeys simple spectral sum rules.
Frequency-integration results in the static spin-spin
correlation function, whereas the fully integrated
weight yields
\begin{eqnarray}
   S(S+1)\, I_{m m'} &\equiv&
   \int\!\tfrac{dk}{2\pi} \!\int\! \tfrac{d\omega}{2\pi} \
   S_{m m'}(k,\omega)
\label{eq:sumrule:gen} \\
&=& S_{m m'}(x_i=0, t=0)
   =
    \bigl\langle \Ssite{0 m}\cdot \Ssite{0 m'}\bigr\rangle
\text{ .}\notag
\end{eqnarray}
Here the prefactor was chosen such that in the present
context of $S=\sfrac{1}{2}$ sites one obtains the normalized
total weight,
\begin{eqnarray}
 I_\mathrm{tot} \equiv
 \sum_{m m'} I_{m' m} = \tfrac{4}{3}
 \bigl\langle
      \Ssite{0}^\mathrm{tot} \cdot
      \Ssite{0}^\mathrm{tot}\bigr\rangle
\equiv \tfrac{4}{3}
 \bigl\langle \Sspin_{0}^2 \bigr\rangle
\gtrsim 1
\text{. }\quad\label{eq:sumrule}
\end{eqnarray}
In the strong rung-coupling regime, where the
local $S=\sfrac{3}{2}$ multiplet is effectively
projected out, we have $I_\mathrm{tot}\simeq1$, which
is assumed in the remainder of this section.
The upper limit given by $I_\mathrm{tot}=5$ holds for
the hypothetical case where the $S=\sfrac{3}{2}$
rung multiplet dominates. In the weak rung-coupling
regime discussed later, we will encounter 
$1\leq I_\mathrm{tot} \lesssim 2$.

In the presence of dimerization, the structure factor
as defined in \Eq{eq:Smm'} becomes complex [while
$S_{m m'}(x,\omega)$ is still real because the ground
state can be taken real for our model, the Fourier
transform in real space becomes complex due to the
broken inversion symmetry]. In this case, we take the
real part of the r.h.s. of \Eq{eq:Smm'} which in the
presence of dimerization is equivalent to
symmetrization of the structure factor w.r.t. the
location of site $i'\in \{0,1\}$. The resulting DSF
then is again symmetric for $k \to -k$, and also
conforms to the standard momentum space definition
and experimentally accessible DSF.

\begin{figure}[tb!]
\begin{center}
\includegraphics[width=1\linewidth]{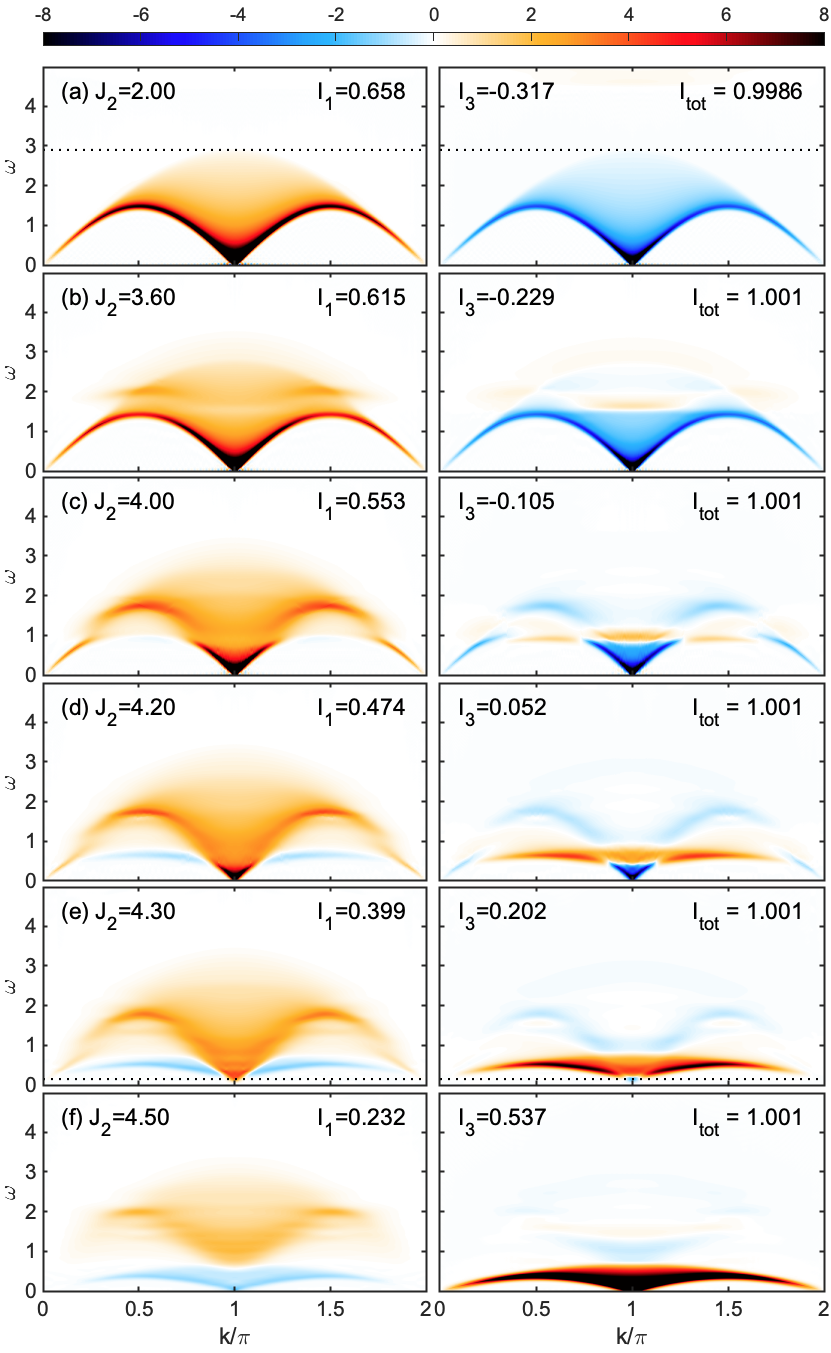}
\end{center}
\caption{
   Dynamical structure factor for the 2-leg ladder
   \eqref{eq:Heff} in the strong rung-coupling regime
   as in \Fig{fig:dim_gap},
   having $J{=}[1,J_2,4]$ for various $J_2$ is indicated
   with the left panels top to bottom. The panel labels (a-f)
   each refer to a row which shares the same $J_2$.
   Left panels
   show $S_1(k,\omega)$, whereas the right panels
   show $S_3(k,\omega)$ with the initial spin
   operator acting on site $m'=1$ or $3$, respectively
   (see text). The corresponding total integrated
   spectral density are also specified ($I_1$ and $I_3$),
   as well as the resulting total $I_\mathrm{tot}
   = 2I_1 + I_3$. The colorbar at the top holds for all
   panels. The spectral data is smoothened with
   $\delta\omega=0.05$ to remove speckles from linear
   prediction, except for $J_2=4.3$, which only uses
   half that value and which also shows a guide for
   the approximate spin gap at $\omega=0.15$ (dotted line).
   Panels (a) show a guide at $\omega=2.9$ which
   approximates the upper bound of the dominant
   spinon band. The weak superimposed wrinkly features
   as in (f) are attributed to state space truncation
   within the DMRG, and hence a numerical artefact.
} \label{fig:DSF1}
\end{figure}

Within DMRG we start from real space,
and hence full real-space resolution.
We explicitly compute
$\bigl\langle \Ssite{i m} (t)
\cdot \Ssite{0 m'}(0) \bigr\rangle
= \langle 0| \Ssite{i m} \cdot \bigl[
  e^{-\iu (H-E_0)}
  \bigl(\Ssite{0 m'} |0\rangle \bigr)\bigr]
$. With $m,m',i'=0$ fixed, the data is computed
from real-time evolution and collected vs. $i$.
For simplicity, we sum the resulting data
over the site index $m$. This 
corresponds to the spectral data at $k_y=0$ w.r.t. $m$,
which is equivalent to using $\Sspin_i$.
The resulting DSF
\begin{eqnarray}
   S_{m'}(k,\omega) \equiv \sum_m S_{m m'}(k,\omega)
\label{eq:Sm'}
\end{eqnarray}
then refers
to the spectral data resulting out of having
acted with the initial spin
operator on rung site $m'$. Since by the
preserved mirror symmetry in the ground state
calculations it follows $S_1 = S_2$,
it suffices to compute $S_1(k,\omega)$ and
$S_3(k,\omega)$ [e.g. as shown in \Fig{fig:DSF1}].
While much of $S_{m'}(k,\omega)$ is dominated
by $m=m'$ which results in a positive spectral
density, it also contains an off-diagonal
contribution $m\neq m'$. Therefore if the
local spin excitation induced at time $t=0$
preferentially propagates to a different rung site
$m\neq m'$, then due to the underlying antiferromagnetic
NN correlation, the spectral density of the DSF
can turn negative for a particular range
in momentum and frequency space.
By properly combining $S_1(k,\omega)$ and
$S_3(k,\omega)$, however, the weighted
average $2S_1(k,\omega) + S_3(k,\omega)$ again
must result in a non-negative spectral density
throughout, as this represents the DSF now
at $k_y=0$ for both, $m$ and $m'$ which
is equivalent to computing the DSF
based on $\bigl\langle \Sspin_{i} (t)
\cdot \Sspin_{0}(0) \bigr\rangle$.
Similarly, the respective total integrated
spectral density is given by $I_\mathrm{tot}
\equiv 2I_1 + I_3 \simeq 1$ [cf. \Eq{eq:sumrule}],
which is well obeyed in the actual numerical
data in the strong rung-coupling regime
[cf. \Fig{fig:DSF1}].

All DSF spectra presented here for the limit of
large rung couplings are computed with the
projected Hamiltonians which have the
$S{=}\sfrac{3}{2}$ rung state space removed,
as this considerably speeds up calculations.
This is justified given that the total weight of the
$S{=}\sfrac{3}{2}$ multiplet states is typically
below 1\% in ground state calculations.
Hence we only expect minor effects as a
result of this simplification here, as verified
in exemplary DSF calculations with the full
rung state space kept (data not shown).
Since the $S{=}\sfrac{3}{2}$ multiplet
lies at high energy from the very outset here,
having $\Delta_0 \gtrsim 5$, this simply means
that faint spinon continua at high energy are
absent, thus only marginally affecting spectral
sum-rules, while at the same time the DSF in the
low-energy regime is well captured. Overall, the
DSF results here are consistent with the
ground-state DMRG analysis above based on the
unprojected Hamiltonians, but greatly compliment
these by adding a dynamical perspective.

Our results for the DSF in the strong coupling regime
for the two-leg ladder \eqref{eq:Heff}
are summarized in \Fig{fig:DSF1}, where left (right)
panels show the DSF $S_{m'}(k,\omega)$ for $m'=1$ ($3$),
respectively. As explained above, the DSF shown
can turn negative, but the combined total DSF,
$2S_1(k,\omega) + S_3(k,\omega)$, is necessarily
positive, throughout, as verified (not shown).
As a demonstration of the latter, we show that the
total spectral density is well-normalized, with
well-obeyed spectral sum rule $I_\mathrm{tot}\simeq1$
to good numerical accuracy, throughout.

The DSF for $J_2=2$ [row \Fig{fig:DSF1}(a)]
shows the behavior of a nearly pristine spin-half
Heisenberg chain. Only a very faint higher-lying band
is visible around $\omega \sim 5$ in $S_3$ 
(right panel). The clearly visible, dominant part
of the spinon continuum is constrained within an upper
bound of $\omega \lesssim 2.9$ (dotted horizontal line)
which is already within 4\% of the expected value
of $(8/9)\pi$ for the limit $1\ll J_2 \ll J_3$.

The very faint higher-lying band around $\omega \sim 5$
in \Fig{fig:DSF1}(a) actually relates to the state
space of the symmetric rung multiplet \Masym.
Having the effective orbital magnetic field
$\alpha-\alpha_0$, its `Zeeman' splitting for
\Fig{fig:DSF1}(a) is $2|\alpha-\alpha_0| \approx 4.6$
which, indeed, coincides with the onset 
of the higher-lying band. Conversely, the low-energy
spinon continuum originates from the
symmetric rung multiplets \Msym.
Now increasing $J_2$ (going to lower rows in
\Fig{fig:DSF1}), the faint \Masym continuum at 
high energies moves downward in energy, 
such that it starts overlapping and interfering
with the low-energy \Msym spinon continuum.

Their different origin also qualitatively translates
into different signs in the DSF $S_3(k,\omega)$,
and hence to different colors in the right panels of
\Fig{fig:DSF1}. There the \Masym spinon
continuum appears positive (reddish), yet
the \Msym spinon continuum appears negative
(blueish).
Hence by following the color coding in the right
panels in \Fig{fig:DSF1} top to bottom, one can
observe with increasing $J_2$ towards $J_3$
and above, that the the original \Masym spinon
continuum at high energies crosses over with
the \Msym spinon continuum at low energies,
which itself then starts lifting off to higher energies.
For the case where the orbital magnetic field is 
approximately zero, e.g. at $J_2=4.3$
in \Fig{fig:DSF1}(e), both spinon sectors show 
a small but finite gap due to dimerization,
with the earlier estimate for the spin gap
$\DeltaS\simeq 0.15$ marked by the vertical
dotted line, and thus consistent with the
dynamical spectral data. This scenario
of {\it crossing state spaces} is also supported
by analyzing ground state entanglement
spectra vs. $J_2$ (not shown).

For the largest $J_2=4.5$ [\Fig{fig:DSF1}],
a different effective spin-half Heisenberg continuum
has developed at low energies (dark red feature
at the bottom right panel). This newly
formed spinon continuum, however, now belongs
to the antisymmetric rung-multiplet, and hence
to the nearly decoupled center spins. Its bandwidth
does not saturate, but will diminish to zero
when $J_2$ is taken to $J_2 \gg J_3$. Since
the leg spins are gapped out, this low-energy
spinon continuum lives predominantly on the center spins. 
Hence $S_3$ is dominated by $m=m'=3$
[cf. \Eq{eq:Sm'}] which is thus expected
positive (reddish in color in \Fig{fig:DSF1}, indeed).

\subsection{Mean-field theory for dimerized phase}

If translational invariance was not spontaneously
broken, then a mean field arguments suggest that
the low lying excitations in the spin sector are
as in the uniform Heisenberg model. It is quantum
critical and hence is susceptible to
perturbations. The most likely relevant operator
is the staggered energy density. This emerges as a
result of spontaneous breaking of the
translational symmetry resulting in dimerization.
In the mean field scheme the staggered energy
density in the spin sector emerges simultaneously
with the staggered component of energy density in
the orbital sector. The spin sector will certainly
lose energy by the dimerization. Therefore one has
to look for a balance to establish whether or not
the dimerized phase gains the overall lower energy.

In order to develop a mean field theory for the dimerized
phase, it is convenient to use Jordan-Wigner transformation
in the orbital ($\tau$) sector. Then with $J_1=1$,
\Eq{eq:Heff} becomes
\begin{eqnarray}
 && \Heff_\alpha
  = \sum_{i} \Bigl(\tfrac{8}{9}\left(
      \mathbf{\Sspin}_{i}\cdot\mathbf{\Sspin}_{i+1}
   \right) \otimes \mathbb{T}_{i,i+1}^{(2)}
  + \iu\alpha \chi_{i}\rho_{i}
  \Bigr),
\label{JW}
\\
&& \mathbb{T}_{i,i+1}^{(2)} = \underbrace{
   \tfrac{1}{4}
{+}\tfrac{\iu}{2} (
      \chi_{i}\rho_{i} {+} \chi_{i+1}\rho_{i+1})
{+}\tfrac{3\iu}{2}\chi_{i}\rho_{i+1}
   }_{\equiv \tilde{\mathbb{T}}_{i,i+1}^{(2)} }
{-}\chi_{i}\rho_{i}\chi_{i+1}\rho_{i+1}
, \notag 
\end{eqnarray}
where
$\chi_{i} {=} \tfrac{1}{\sqrt{2}} (c_i {+} c_i^\dagger)$ and
$\rho_{i} {=} \tfrac{\iu}{\sqrt{2}} (c_i^\dagger {-} c_i)$
are Majorana and thus real fermions that are subject to the
anticommutation relations
$\{\chi_{i},\chi_j\} = \{\rho_{i},\rho_j\} = \delta_{i j}$. 
Due to the reality of the Majorana fermions,
Hamiltonian \eqref{eq:Heff} is Hermitian as it stands,
yet may be symmetrized via $\Heff_\alpha {=}
\tfrac{1}{2}( \Heff_\alpha {+} \Heff_\alpha\!\,^\dagger)$.
To simplify matters we will omit the four-fermion (last) term
above that corresponds to the $\Torb^z\Torb^z$ term
in \Eq{eq:Heff:2}, as we do not aim for precision here,
leaving this to the numerical calculations.
This results in the mean field
approximation of \Eq{JW},
\begin{subequations}\label{eq:H:MF-dimer}
\begin{eqnarray}
   H^\mathrm{MF} &\equiv& \sum_{i}
      \beta_i
      \left(\mathbf{\Sspin}_{i}\cdot\mathbf{\Sspin}_{i+1}\right)
   {+} \underbrace{\iu \sum_{i} 
     \chi_{i}
     \bigl(-\tfrac{3\alpha_i}{2}
     \rho_{i+1} {+} h \rho_{i}\bigr)}_{\equiv H_\tau}
\quad\quad \label{eq:Htau:MF-2} \\[-3.2ex]\notag
\end{eqnarray}
having
\begin{align}
 \alpha_i
    &\equiv -\tfrac{8}{9} \langle
     \mathbf{\Sspin}_{i}\cdot\mathbf{\Sspin}_{i+1} \rangle
    &\equiv& \alpha_0 \, [1+ \delta(-1)^i]
 \label{eq:alpha:i}\\
 \beta_i
    &\equiv \phantom{+} \tfrac{8}{9} \bigl\langle
       \tilde{\mathbb{T}}_{i,i+1}^{(2)} 
    \bigr\rangle 
    &\equiv& \beta_0 \, [1+ \gamma(-1)^i]
 \label{eq:beta:i}\\
    h &\equiv \alpha
     - \tfrac{1}{2}(\alpha_{i-1} + \alpha_i)
     &=& \alpha - \alpha_0
 \label{eq:h}
\end{align}
\end{subequations}
with $\alpha_0\approx 0.394$ [cf. \Eq{eq:alpha0}].
Here $\delta$ and $\gamma$ are additional parameters
to describe the strength of dimerization in the
spin and orbital sector, respectively.

Further progress can be made assuming that the
resulting spectral gap is small in comparison with
the band width which, as we will see, is consistent
with the numerical calculations. Under this assumption
we can bosonize the spin part of (\ref{JW}).
The uniform part of the Heisenberg Hamiltonian
becomes the Gaussian model and the staggered part is
$(-1)^i\left(\mathbf{\Sspin}_{i}\cdot\mathbf{\Sspin}_{n+1}\right)
= A\cos(\sqrt{2\pi}\Phi)$ where one can conclude
from Ref.\,\onlinecite{Lukyanov98} that coefficient $A \sim 1$. 
Then we obtain the following sine-Gordon Lagrangian:
\begin{eqnarray}
    L^\mathrm{MF}_s {=}
      \int dx\Big[\tfrac{1}{2v}(\partial_{\tau}\Phi)^2 
    + \tfrac{v}{2}(\partial_x\Phi)^2
    - \tfrac{m_0^2}{2\pi}\cos(\sqrt{2\pi}\Phi)\Big]
\text{, }\quad\label{SG}
\end{eqnarray}
where $v = \beta_0 \pi$ and $m_0^2 = 2\pi A\gamma$.
This sine-Gordon model has a hidden SU(2) symmetry. Its
excitations are massive and consist of one massive
triplet (soliton, antisoliton and the first breather)
with mass $m_t \approx 0.893 \, m_0^{4/3} $ as can be
extracted from \cite{Lukyanov97}, and the second
breather with mass $\sqrt 3\, m_t$. Then we have
\begin{eqnarray}
    \delta \sim \langle\cos(\sqrt{2\pi}\Phi)\rangle
    \approx 0.163 \, \sqrt{m_t}
 =  0.154 \, (A\gamma)^{1/3}
\text{ .}\quad\label{eq:delta:gamma}
\end{eqnarray}
Next we diagonalize the $\tau$-part of the Hamiltonian
where we also aim to obtain a relation between
the dimerization parameters $\delta$ and $\gamma$.
In momentum space with a 2-site unit cell,
the Hamiltonian assumes the matrix form,
\begin{eqnarray}
&& H_{\tau}= \sum_{k>0}\Psi^\dagger(k)H_{\tau}(k)\, \Psi(k), \\
&& H_{\tau}(k) =
 \tfrac{3 \iu \alpha_0}{4}
 \begin{pmatrix}
    0 & \tilde h & 0 & (1+\delta) \\
    -\tilde h & 0 & \hspace{-.1in} (1-\delta)e^{-ik} & 0\\
    0 & \hspace{-.1in}-(1-\delta)e^{ik} & 0 & \tilde h\\
      -(1+\delta) \hspace{-.1in} & 0 & -\tilde h & 0
  \end{pmatrix}
\notag
\end{eqnarray}
where $h{\equiv} \tfrac{3\alpha_0}{2} \tilde h$ and
$\Psi^T{=}((\chi,\rho)_{A,k}(\chi,\rho)_{B,k})$.
Its eigenvalues $\varepsilon(k) \equiv
\tfrac{3\alpha_0}{4} \tilde{\varepsilon}(k)$ out of
$\mathrm{det}(H_{\tau} - \varepsilon)=0$ are given by
\begin{eqnarray}
  \tilde{\varepsilon}^2
   &=& 1 + \delta^2 + \tilde{h}^2 \pm 2\delta_k
\text{ ,}\label{eq:epsilon:k}
\end{eqnarray}
with $ \delta_k^2 \equiv
\delta^2 + \tilde{h}^2 \bigl[
   1 - (1-\delta^2) \sin^2 (\tfrac{k}{2})
\bigr]$.
The dimerization $\delta$
shifts the critical field and renormalizes the velocity,
as seen by expanding around small $k$,
\begin{eqnarray}
 \tilde{\varepsilon}^2
 \simeq \Bigl(
   1 \pm \sqrt{\delta^2 + \tilde{h}^2}\ 
 \Bigr)^2
 \mp \tfrac{ (1-\delta^2)\tilde{h}^2}{
    4\sqrt{\delta^2+\tilde{h}^2}}\, k^2
\end{eqnarray}
Now by making use of the Hellmann-Feynman theorem,
we also have from \Eqs{eq:H:MF-dimer} above,
\begin{eqnarray}
  \tfrac{\partial}{\partial\delta} \langle H_\tau \rangle
&=&
  \langle \tfrac{\partial H_\tau}{\partial\delta} \rangle
= - \alpha_0 \sum_i (-1)^i \tfrac{3\iu}{2} 
    \langle \chi_i \rho_{i+1} \rangle
\notag \\
&=& - \alpha_0 \, \tfrac{N}{2} \underbrace{\bigl\langle
     \tilde{\mathbb{T}}_{2,3}^{(2)}
   - \tilde{\mathbb{T}}_{1,2}^{(2)} 
  \bigr\rangle}_{=\tfrac{9}{8} (\beta_2 - \beta_1)}
 = - \tfrac{9N}{8} \alpha_0\beta_0 \,\gamma
\text{ .}\label{eq:dHtau}
\end{eqnarray}
Here in the orbital sector, $\delta$ is considered
an external parameter that gives rise to a finite
orbital dimerization $\gamma$.
\begin{figure}[tb!]
\begin{center}
\includegraphics[width=1\linewidth]{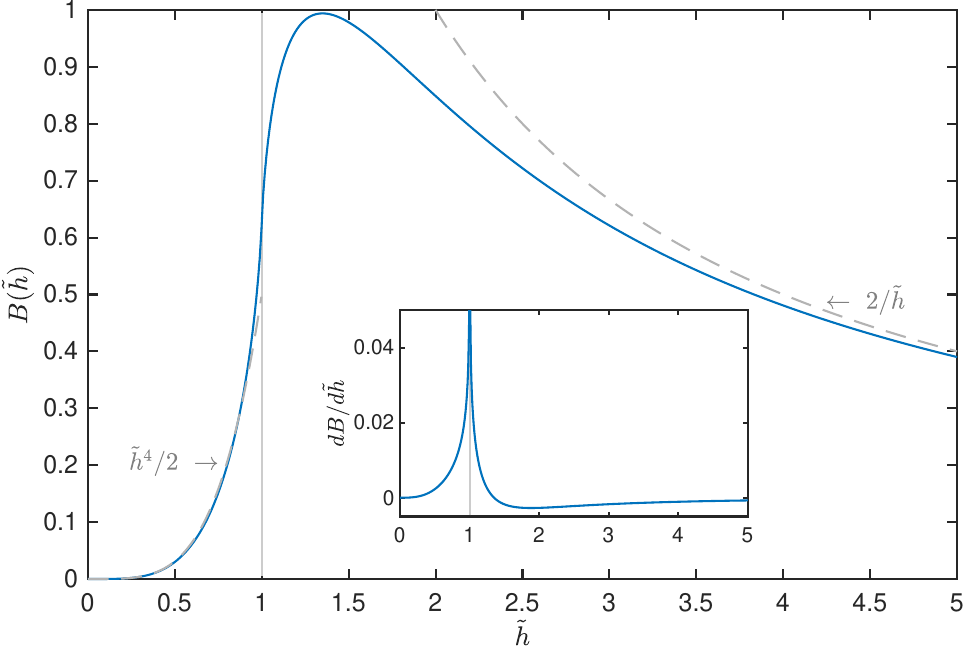}
\end{center}
\caption{
  The dimensionless
  function $B(\tilde h)$ from \Eq{Bfunc}.
  Dashed gray lines indicate asymptotic behavior.
  The inset shows a weak non-analyticity at
  $\tilde{h}=1$ resulting in a vertical slope in main panel.
} \label{fig:B}
\end{figure}
Therefore,
\begin{subequations}
\begin{eqnarray}
   \gamma &=& \tfrac{8}{9\alpha_0\beta_0} \Bigl(
   -\tfrac{1}{N}
   \tfrac{\partial}{\partial\delta} \langle H_\tau\rangle
   \Bigr)
\notag \\
   &=& \tfrac{1}{9\alpha_0\beta_0\pi} \tfrac{\partial}{\partial\delta}
   \int\limits_{-\infty}^\infty \!\! d\epsilon
   \int\limits_{0}^\pi \!\! \tfrac{dk}{2\pi}
   \ln\det \bigl[H_\tau(k) - i\epsilon\bigr]
\notag \\
&\equiv& \tfrac{B(\tilde{h})}{6\beta_0} \,\delta
\end{eqnarray}
where
\begin{eqnarray}
B(\tilde{h}) &\equiv &
  \int\limits_0^{\pi} \tfrac{2 dk}{\pi} 
  \sum_{\sigma=\pm 1}
  \tfrac{1+ \sigma \tfrac{
    1+{\tilde h}^2\sin^2(k/2)}{
      {\tilde h}|\cos(k/2)|}}
  {\bigl(1+{\tilde h}^2 + 2\sigma\tilde h|\cos \tfrac{k}{2}|\bigr)^{1/2}} \ \ge 0
\text{ ,}\quad\label{Bfunc}
\end{eqnarray}
\end{subequations}
where we expanded to linear order around $\delta{=}0$.
With \Eq{eq:epsilon:k}, $\langle H_\tau\rangle$ is an
even function in $\delta$. Therefore,
$\gamma(\delta)$ is necessarily odd and hence, to
lowest order, linear in $\delta$. Therefore it also
holds $B(\tilde h) \propto -\tfrac{\partial^2
}{\partial\delta^2}\langle H_\tau\rangle$. With $h$
acting like an external magnetic field in the orbital
sector, \Eq{Bfunc} yields a linear relationship
between the dimerization parameters $\gamma$ and
$\delta$. Matching this with the earlier relation in
\Eq{eq:delta:gamma}, $\gamma\propto \delta^3$, we get
two solutions, (i) the non-dimerized phase at
$\gamma=\delta=0$, as well as (ii) the non-trivial
dimerized solution where with $(A\gamma)^{2/3} =
A\gamma / (A\gamma)^{1/3}\simeq 0.154\, A \  % \cdot
\gamma/\delta$, i.e.,
\begin{eqnarray}
    (A\gamma)^{2/3} = 0.154\, \tfrac{A}{6\beta_0}
    B(\tfrac{2h}{3\alpha_0})
\end{eqnarray}
As seen in the numerical evaluation of the function
$B(\tilde{h})$ in \Fig{fig:B}, it vanishes quartically
at $\tilde{h}=0$, has a non-analyticity with vertical
slope at $\tilde{h}=1$, followed by a maximum at
$\tilde{h}_c \sim 1.35$, and then for large
$\tilde{h}$ decays like $2/\tilde{h}$. Hence there is
an area of the phase diagram where the assumption
$\gamma \ll 1$ is valid, and so the current
calculation is self-consistent.

Having $B(\tilde{h}) \sim - \tfrac{1}{\delta}
\tfrac{\partial}{\partial\delta} \langle H_\tau\rangle > 0$,
it follows that an orbital dimerization pattern that
is aligned with the dimerization in the spin sector
(in the sense that $\gamma$ and $\delta$ have the same
sign), this allows the orbital sector to {\it lower}
its energy. Hence we conclude that the system favors
dimerization and in the strong coupling limit a
self-consistent dimerized solution always exists, at
least for these somewhat simplified calculations with
the $\Torb^z\Torb^z$ term omitted. This conclusion is
consistent with our DMRG data which shows a noticeable
dimerization in the vicinity of $|h| \ll 1$ [e.g. see
\Fig{fig:snapshots}].

\section{The limit of weak rung exchange}
\label{seq:wk-rung-cpl}

In the limit of weak rung couplings, the full state
space of the rungs needs to be included.
Specifically, the $S=\sfrac{3}{2}$ symmetry sector can
no longer be simply integrated out. In this section,
we start with the theoretical description, followed by
DMRG simulations of the dynamical structure factor.
The results are mutually consistent. In contrast to
the strong rung-coupling regime above, we do not find
any indication for dimerization here. Instead, we
find a low-energy coherent branch. Consequently,
there needs to be a quantum phase transition when
decreasing $J_2,J_3 \gg 1$ to small values $J_2,J_3 <
1$, the precise determination of which is left for
future studies. By comparison, it may be noted that a
fermionic model on the same lattice as in
\Fig{fig:model} in the weak rung-coupling regime also
features flat bands that are predominantly associated
with the weakly coupled center spins.

\subsection{Field theoretic approach}

If the interchain exchange interactions are small,
$J_2,J_3 \ll J_1=1$ we can use the continuum limit
description. In this limit the chains are described by
the critical SU$_1$(2) Wess-Zumino-Novikov-Witten (WZNW)
theories \cite{affleck,Tsvelik_book} and the
interchain interaction and the interaction with the
central spins are perturbations to this critical
model. Both perturbations are relevant, but the
interaction with the central spins is more relevant
since it has scaling dimension 1/2 and the interchain
coupling of the staggered magnetizations has dimension
1. We will consider the case when the interchain
exchange is zero first.
 
Our derivation is a strict generalization of the one
for a single chain coupled to dangling spins presented
in \cite{Igor}. We will reproduce it below with the
appropriate modifications. It is the most convenient
to combine the path integral representation for the
middle spins with the field theory description for
the legs. In this representation, the middle spins
are replaced as $\Ssite{0,j} = S_0{\bf N}_j$, where
${\bf N}_j$ is a unit vector field with the Berry
phase action. In the current context
$S_0{=}\sfrac{1}{2}$, but we prefer to keep it
arbitrary for the time being. As far as the
Heisenberg chains are concerned, at energies $\ll 1$
we can use the field theory description, which is
given by the SU$_1$(2) WZNW theory. The
resulting action for energies $\ll 1$ is given by,
\bea
S &=& \sum_j S_0 A[{\bf N}_j] + W[g_1] + W[g_2] 
\notag\\
&& + \iu\gamma \sum_j (-1)^j\sum_{a=1,2}\int d \tau 
     {\bf N}_j \mathrm{Tr}[\vec\s(g_a^+-g_a)]
,\quad\label{action}
\eea
where ${\bf S}_0 = S_0{\bf N}, ~~ {\bf N}^2=1$, $g_a(\tau,x)$
are the SU(2) matrix fields, and $W[g]$ is the action of the
SU$_1$(2) WZNW theory, $A[{\bf N}]$ is the Berry phase and
$\gamma \sim S_0 J_3$.
The Heisenberg spins are related to the WZNW fields,
\bea
   \Ssite{j,a} = \tfrac{\iu}{2\pi}
   \mathrm{Tr}({\vec\s}g_a\p_x g_a^+) + \iu(-1)^jC\mathrm{Tr}[\vec\s(g_a-g_a^+)]
,\quad\label{bozonspin}
\eea
where $C$ is a nonuniversal amplitude. The WZNW model is a
critical theory with a linear excitation spectrum,
$\omega = v|k|, ~~v = \pi J/2$.

In the interaction term in (\ref{action}) we kept only
the most relevant term, which describes the
interaction of the central spins with the staggered
magnetization of the Heisenberg chains. This action
is not yet what we need since the central spin
variables remain lattice ones. In order to obtain the
continuum limit, we have to integrate out the fast
components of the central spins. We assume that at low
energies these spins have a short range
antiferromagnetic order, so we can write,
\bea
   {\bf N}_j = 
   {\bf m}(x) + (-1)^j(1-{\bf m}^2)^{1/2}{\bf n}(x), ~~ x=a_0 j
,\quad
\eea
where ${\bf n}^2 =1$ and $|m| \ll 1$. The validity of this
assumption is justified by the final result which demonstrates
that the correlation length of the middle spins is much larger
than the lattice constant. Substituting this into
(\ref{action}) and following the well known procedure
\cite{Haldane_PL_83,Tsvelik_book}, we obtain
\bea
   S &=& \int \rd\tau \rd x\Big\{ \tfrac{\iu S_0}{2}\bigl({\bf n}[\p_{\tau}{\bf n}\times\p_x{\bf n}]\bigr) 
   + \iu S_0({\bf m}[{\bf n}\times\p_{\tau}{\bf n}])
\nonumber \\
   && +\iu\gamma (1-{\bf m}^2)^{1/2}\ \mathrm{Tr}\sum_a[(\vec\s{\bf n})(g_a-g_a^+)]\Big\} \notag\\
   && + W[g_1] +W[g_2].
\label{S2}
\eea
Now notice that $G = \iu(\vec\s{\bf n})$ is an SU(2)
matrix. Hence, $h_a= g_aG^+$ is also an SU(2) matrix
and we can use the identity \cite{polyakov},
\bea
W[hG] = W[h] +W[G] +
  \int\tfrac{\rd\tau \rd x }{2\pi}\,
  \mathrm{Tr}(h^+\p h G\bar\p G^+)
 \qquad \label{identity}
\eea
with $\p,\bar\p = \tfrac{1}{2}(\p_{\tau} \mp \iu v\p_x)$,
so that the action (\ref{S2}) becomes
\begin{eqnarray}
&& S = S_\mathrm{mass} {+} S_{m} {+} S_{n} + \sum_a
  \int\tfrac{\rd\tau \rd x }{2\pi}\,
  \mathrm{Tr}(h_a^+\p h_a G\bar\p G^+) \qquad
\end{eqnarray}
where
\begin{eqnarray}
   S_\mathrm{mass} &=&
   W[h_1] + W[h_2] + \gamma \sum_a\int \rd\tau \rd x \, \mathrm{Tr}(h_a+h_a^+)
\qquad\label{h}\\[-1.4ex]
   S_m &=& \int \rd\tau \rd x \ \bigl\{ \tfrac{D}{2}{\bf m}^2
   + \iu S_0 ({\bf m}[{\bf n}\times\p_{\tau}{\bf n}])\bigr\}
\label{Sm}\\
   S_n &=& 2W[\iu(\vec\s{\bf n})] +S_0 (\text{top-term})
\label{Sn}\\
   S_\mathrm{top} &=& \int \rd\tau \rd x \,
  \tfrac{\iu }{2}\bigl({\bf n}[\p_{\tau}{\bf n}\times\p_x{\bf n}]\bigr) ,
\end{eqnarray}
having
\be
   D = \gamma \sum_a\la \mathrm{Tr}(h_a+h_a^+)\ra \sim \gamma^{4/3}.
\label{D}
\ee
The latter estimate follows from the fact that the $h$-matrix
operator in the SU$_1$(2) WZNW model has scaling
dimension 1/2. In a (1+1)-dimensional critical
theory, a relevant perturbation with a scaling
dimension $d$ and coupling constant $\lambda$
generates a spectral gap, $\Lambda \sim \lambda^{1/(2-d)}$.
Consequently, the perturbation itself acquires a
vacuum expectation value, $\sim \Lambda^d \sim
\lambda^{d/(2-d)}$, giving rise to (\ref{D}).

Integrating over ${\bf m}$ and taking into account that
\bea
  W[\iu(\vec\s{\bf n})] &=& 
    \tfrac{1}{2\pi}\int \rd\tau \rd x[v^{-1}(\p_{\tau}{\bf n})^2 +v(\p_x{\bf n})^2]
  + \tfrac{1}{2}
  S_\mathrm{top},\notag\\ \label{wztop}
\eea
we obtain the effective Lagrangian density for the slow field ${\bf n}$:
\bea
{\cal L} &=& \tfrac{1}{2}\bigl(
    \tfrac{S_0^2}{D} + \tfrac{1}{\pi v}
  \bigr)(\p_{\tau}{\bf n})^2 + \tfrac{v}{2\pi}(\p_x{\bf n})^2
+ \tfrac{\iu S_0}{2}\bigl(
    {\bf n}[\p_{\tau}{\bf n}\times\p_x{\bf n}]
  \bigr) \notag\\ % ,
\label{sigma}
\eea
plus the action for the massive part for each $a=1,2$:
\bea
  S_\mathrm{mass} &=& W[h] +\gamma \int \rd\tau \rd x \mathrm{Tr}(h+h^+)
\nonumber\\
   &+& \int \rd\tau \rd x \mathrm{Tr}({\bf J}_L[{\bf n}\times\bar\p {\bf n}]).
\label{mass1}
\eea
This theory without the last term is, in fact,
equivalent to the famous sine-Gordon model at the
special value of the coupling constant $\beta^2 =
2\pi$. Indeed, the SU$_1$(2) WZNW model is equivalent
to the Gaussian theory and Tr$(h+h^+) \sim
\cos(\sqrt{2\pi}\phi)$ such that 
\bea
   W[h] &+&\gamma \int \rd\tau \rd x \mathrm{Tr}(h+h^+) \nonumber\\
   &=& \int \rd\tau \rd x\Big[\tfrac{1}{2}(\p_{\mu}\phi)^2
     - \tilde\gamma\cos(\sqrt{2\pi}\phi)\Big].
\eea
This theory is massive and the spectrum consists of an
SU(2) triplet with mass $M \sim \gamma^{2/3}$ composed
of sine-Gordon kink and antikink excitations and the first
breather, and the second breather with mass $\sqrt 3 M$. 

Note that the contribution to the topological term
from (\ref{wztop}) shifts the coefficient by one which
is equivalent to zero. The mass gap $\Lambda$ serves
as the ultraviolet cut-off for the sigma model
(\ref{sigma}). The corrections to the sigma model
generated by the last term in (\ref{mass1}) carry a higher
power of gradients of the ${\bf n}$-field and therefore
can be discarded for momenta $< \Lambda v^{-1} $.

For the case relevant to this paper, the
$S_0{=}\sfrac{1}{2}$ sigma model (\ref{sigma}) has a
gapless spectrum in the same universality class as the
$S{=}\sfrac{1}{2}$ Heisenberg chain \cite{zamfat}.
This mode is slow since the corresponding velocity is 
\bea
    c^2 = \tfrac{v^2}{1 + \frac{\pi v}{4D}}.
\label{eq:WCR:c2}
\eea
We emphasize that the above treatment is valid only in
the region of energies much smaller than the
excitation bandwidth. As is evident from the DMRG
calculations, indeed, for most of the Brillouin zone
the spectrum of the gapless mode is rather flat which
is consistent with the smallness of the velocity
(\ref{eq:WCR:c2}). The linear spectrum holds only in
the vicinity of zero or $\pi$ wave vectors.
On the other hand models describing rotated spins
(\ref{mass1}) have a spectrum with a gap $\Delta_3
\cong J_3^{2/3}$ (all energies in units $J_1=1$). 
 
\begin{figure}[tb!]
\begin{center}
\includegraphics[width=1\linewidth]{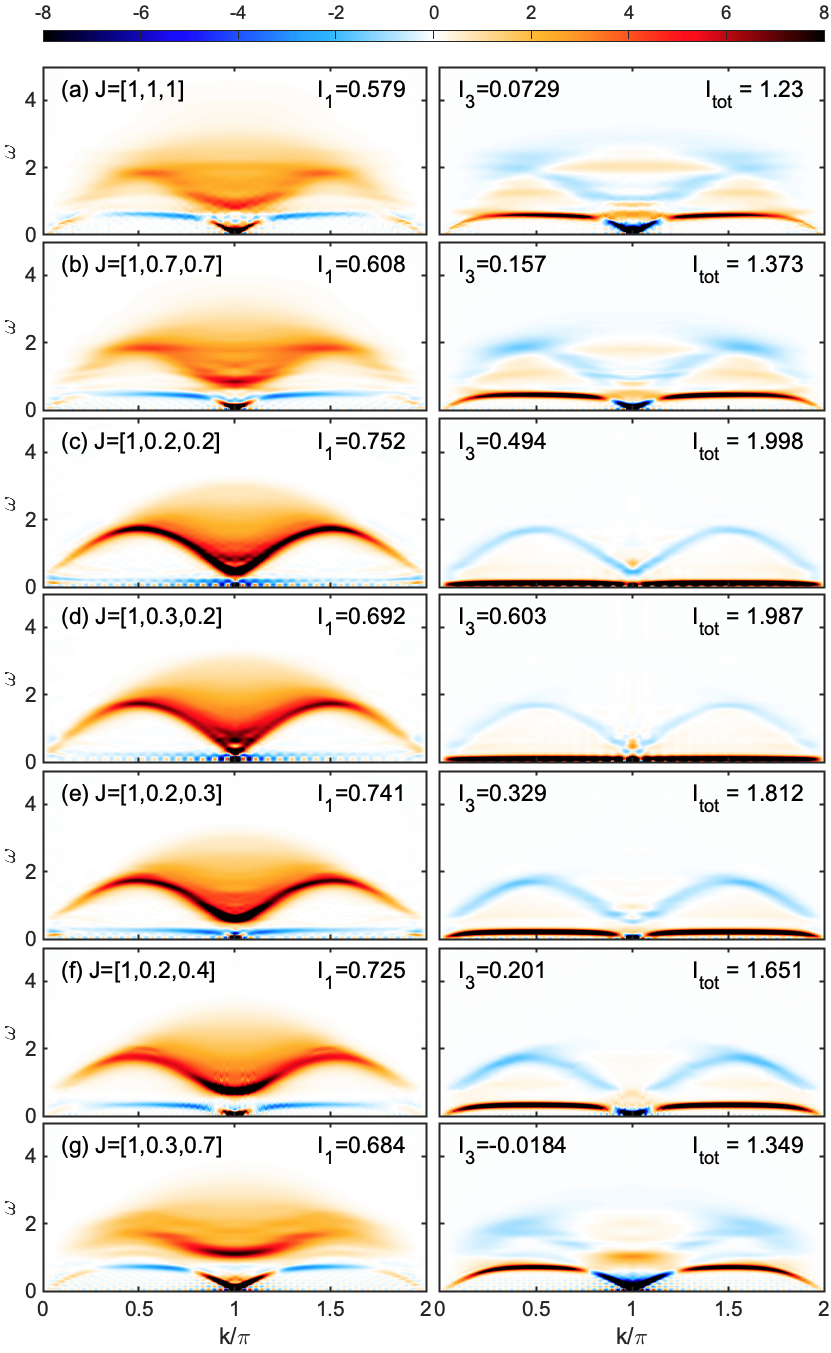}
\end{center}
\caption{
   Dynamical structure factor obtained via DMRG
   for the 2-leg ladder \eqref{fig:model}
   in the intermediate to weak rung-coupling regime
   for $L{=}64$ rungs [except for row (f) which has
   $L{=}128$]. The coupling
   strength is specified with the panels (panel label
   and $J$ hold per row).
   Rows (a-c) have decreasing isotropic rung coupling
   $J_2=J_3$. Rows (c-e) have $J_2<J_3\ll 1$
   with $\alpha=0$, $\alpha<0$, and $\alpha>0$,
   respectively. The remainder of the rows have $\alpha>0$
   then again with increasing rung coupling. 
   Exactly same analysis as in \Fig{fig:DSF1} otherwise.
   In the present case, however,
   $I_\mathrm{tot}>1$ indicates that there is also
   a significant admixture of the $S=\sfrac{3}{2}$
   rung multiplet.
} \label{fig:DSF2}
\end{figure}

The spectral weight of the slow gapless mode is
concentrated on the central spins which is fully
consistent with the results of the DMRG calculations
displayed in \Fig{fig:DSF2}. As for the spins located
on the legs, they receive only a portion of it. The
spin-spin correlation functions of spins located on
the legs of the ladder are symmetric, and thus also in
a phase with unbroken \Ztwo symmetry. Substituting the
expression for $g = G^+h$ into (\ref{bozonspin})
we get for the staggered magnetization,
\bea
 && {\bf S}_\mathrm{stag}
     \sim {\bf n}\la\cos(\sqrt{2\pi}\phi)\ra + [{\bf n}\times{\bf K}]
\label{newspin}\\
 && {\bf K} = \bigl(
     \sin(\sqrt{2\pi}\phi)  ,\, 
     \cos(\sqrt{2\pi}\theta),\, 
     \sin(\sqrt{2\pi}\theta)
   \bigr)
\eea
where $\theta$ is the field dual to $\phi$. The
correlation functions of the sine-Gordon model are
well known, in particular, for this value of $\beta$
the lowest part of the spectral weight consists of a
coherent peak. As we can see from (\ref{newspin}) in
the spectral weight of the leg spins this peak will be
broadened by the emission of soft excitations of the
${\bf n}$-field. Such broadening cannot exceed the
bandwidth of these excitations. Such a picture is
consistent with \Fig{fig:DSF2}.

The solution presented above is valid when the
spectral gap of the ``rotated'' fields $\Delta_3 \sim
J_3^{2/3}$ is much larger than the spectral gap
generated by the direct interchain exchange,
$\Delta_2\sim J_2$, i.e., $1 \gg J_3^{2/3} \gg J_2$
(all energies in units $J_1=1$), and holds only in the
vicinity of the wave vectors $0$ and $\pi$. where
excitations of the {\bf n} field are gapless in
agreement with the DMRG. Matter of fact, the opposite
case, $\Delta_3 \ll \Delta_2$, would not qualitatively
differ from this one. Indeed, the strong interchain
coupling would generate a spectral gap in the
spin-half ladder. Integrating out the gapped mode we
would get an effective exchange interaction between
the central spins. These spins then would form a spin
$S{=}\sfrac{1}{2}$ Heisenberg chain with gapless
excitations.
In both limits considered above the spin-spin
correlation functions of spins located on the legs of
the ladder are symmetric. From a topological
perspective, the weak coupling limit thus is also
trivial with no hidden order.

\begin{figure}[tb!]
\begin{center}
\includegraphics[width=1\linewidth]{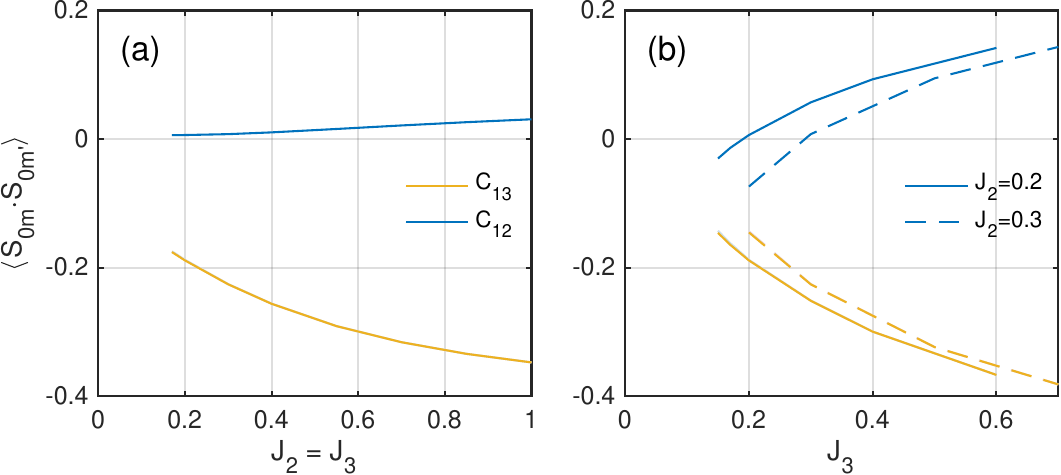}
\end{center}
\caption{ Static spin expectation values $C_{m m'} 
   \equiv \langle {\bf S}_{i m} \cdot {\bf S}_{i m'}\rangle$
   within the same rung $i=0$ in the system center of
   $L=64$ ladders with open BC for (a) $(J_2{=}J_3)
   \le J_1=1$, and (b) vs. $J_3$ for fixed smaller
   $J_2 \ll 1$ as indicated in the legend. The color
   coding in the legend in (a) holds for both panels.
} \label{fig:BaIrO-wk}
\end{figure}

\subsection{Numerical analysis}

In the weak rung-coupling regime, the legs of the
ladder in the model system \eqref{eq:model} tend to be
weakly coupled from a static perspective. This is
demonstrated via the static spin-spin correlators
$C_{m m'}^{(i)} {\equiv} \langle {\bf S}_{i m} \cdot
{\bf S}_{i m'}\rangle$ between the sites of the same
rung $i=0$ in the system center in \Fig{fig:BaIrO-wk}.
For $J_2=J_3$ [\Fig{fig:BaIrO-wk}(a)], the direct
leg-spin correlation, $C_{12}$ (blue line), diminishes
much faster than the correlation of the leg-spins to
the center spin (yellow line, same for both legs).
For fixed small but finite $J_2$, tuning the coupling
$J_3$ [\Fig{fig:BaIrO-wk}(b)] induces a sign change
of the direct leg correlation $C_{12}$. Eventually, it
saturates to a finite negative value for $J_3\to 0$
since \mbox{$J_2>0$}. At the same time, the correlation
$C_{13}$ to the center spin (yellow line; same for
both legs) needs to vanish there. Hence the lines in
\Fig{fig:BaIrO-wk}(b) eventually cross for
sufficiently small $J_3$.

The dynamical behavior in the weak rung-coupling
regime is summarized in the DSF simulations presented
in \Fig{fig:DSF2}. These calculations are
considerably more demanding numerically, since the
full rung state space needs to be included. Clearly,
for $J_2,J_3 \lesssim 1$ the $S{=}\sfrac{3}{2}$ rung
states are also expected to pick up considerable
weight, and hence cannot be ignored. This is
explicitly seen in \Fig{fig:DSF2} by having
$I_\mathrm{tot}>1$, with the total weight in the $S =
\sfrac{3}{2}$ given by $(I_\mathrm{tot}{-}1)/4$
[cf.~\Eq{eq:sumrule}]. Therefore the simulations
here are constrained to shorter ladders of length
$L=64$, except for row \Fig{fig:DSF2}(f) which has
$L=128$. The affordable time range prior to Fourier
transform is also more constrained which translates
into less overall spectral resolution. With the help
of linear prediction we can sharpen certain physical
features in the DSF, yet at the price of also other
`wrinkly' artificial features. Nevertheless, we
preferred smaller subsequent broadening (same as in
\Fig{fig:DSF1}) over significant over-broadening of
the data to completely smear out artifacts due to
DMRG truncation. Also due to the shorter system size,
discretization artifacts are also seen vs. momentum
$k$. Bearing this in mind, we proceed to the physical
interpretation of the results. 

Within our energy resolution, all spectra are gapless.
The low-energy regime of the DSF is dominated by a
sharp coherent branch below the spinon continuum. Its
energy quickly diminishes with decreasing $J_2 \sim
J_3 <1$, and develops a close to flat dispersion over
an extended momentum range [e.g., see center rows in
\Fig{fig:DSF1}]. It is much sharper in energy, and
does not show the energy spread typically seen with
spinon continua. In this sense, the weak rung-coupling
regime is qualitatively different from the crossings
of the two spinon continua that was observed in the
strong rung-coupling regime in \Fig{fig:DSF1}.
Similar to the strong rung-coupling regime, however,
the low-energy branch here is also largely associated
with the center spins, as implied by the sign (color)
in the spectral data in left vs. right panels in
\Fig{fig:DSF2}. Specifically, we see a fainter
negative (blue), yet a strong positive (dark red)
spectral weight in the low energy branch in the right
panels, which relates to off-diagonal ($m'\neq 3$) vs.
diagonal ($m'=3$) correlations, respectively.

The very flat branch close to zero energy as seen for
$J_2,J_3 \sim 0.2$ in the middle panels of
\Fig{fig:DSF2}(c-e) nearly resembles static scatterers.
Due to frustration, and the spectral data above, it
can be argued that this is due to nearly decoupled
center spins. Conversely then, from an experimental
point of view, this coherent low-energy branch may be
very difficult to distinguish from the static
background that arise from actual impurities and
imperfections in samples. In combination, it may also
give rise to spin freezing \cite{Zaliznyak99} with
reference to the magnetic moments on the center sites.
Overall, the numerical results presented here are in
qualitative agreement with the analytical discussion
of the weak rung-coupling regime above.

\section{Conclusions}
\label{sec:conclusions}

We have studied the model of a spin $S{=}\sfrac{1}{2}$
Heisenberg ladder with trimer rungs in the
antiferromagnetic regime. The two legs of the ladder
are coupled by a direct exchange, yet also indirectly,
via an additional center spin for each rung which
introduces frustration. Many results are consistent
with the general expectations. In particular, there is
a significant part of the phase diagram where the
spectrum of the spin excitations is gapless and
critical belonging to the universality class of the
spin $S{=}\sfrac{1}{2}$ Heisenberg antiferromagnet.
The novel feature is the presence of dimerization in
the regime of strong rung coupling.
For reference, the model studied can be considered as
a version of a three-leg ladder with anisotropic rung
coupling and where the coupling along the third leg is
taken to zero. We numerically show that the dimerized
phase in our model smoothly connects to the dimerized
phase that has been previously reported on the
isotropic three-leg ladder \cite{Nishimoto09}. This
provides support and further physical insight into our
findings, namely that the dimerization is driven by a
frustration-driven spin-Peierls transition
\cite{Kawano97}.

In the regime of weak rung-coupling, we find a sharp
coherent low-energy branch. It is largely associated
with the center spins which become nearly decoupled.
This is consistent with the experimental observation
in the trimer magnet Ba$_4$Ir$_3$O$_{10}$ we started
out from where the onset of AF ordering is deferred to
extremely low temperatures as compared to the
estimated exchange energies \cite{Cao20}. Note that
when returning to the 2D hexagonal model system in
\Fig{fig:model}(b), the center spins in our quasi-1D
reduction form an effective square lattice where
N\'eel order eventually may be expected.

\begin{acknowledgments}

Brookhaven National Laboratory was supported by U.S. Department
of Energy (DOE) Office of Basic Energy Sciences (BES), Division
of Materials Sciences and Engineering.
% under contract No. DE-SC0012704 (no longer required; known automatically)

\end{acknowledgments}

\bibliography{Ising_quantum_2}

\appendix
\mbox{\ }\newpage\clearpage

  \def\u{\uparrow}
  \def\d{\downarrow}

\section{Matrix representation of many-body downfolding}
\label{App:downfolding}

Here we present a general method of many-body
downfolding in matrix representation. We show the
method by means of the specific example for the model
\Eq{eq:model}. The basis set is given by the $2^3{=}8$
states $|S^z_1 S^z_2 S^z_3\rangle \equiv |S^z_1\rangle
|S^z_2\rangle |S^z_3\rangle $ which form the direct
product space of $S_1 \otimes S_2 \otimes S_3$, where
$S^z_m \in \{\uparrow,\downarrow\}$ denote spin up and
down in a spin-half state. The orthonormal
eigenvectors of $H_i^\mathrm{rung}$ are given by
\begin{eqnarray}
   |\Psi_1\rangle &\equiv& |\tfrac{1}{2},\tfrac{+1}{2}\rangle^+ =
   \tfrac{1}{\sqrt{6}}\bigl[\phantom{+}|(\u\d + \d\u)\u\rangle
     - 2 |{\u\u\d}\rangle \bigr] \notag \\
   |\Psi_2\rangle &\equiv& |\tfrac{1}{2},\tfrac{-1}{2}\rangle^+ =
   \tfrac{1}{\sqrt{6}}\bigl[-|(\d\u + \u\d) \d\rangle
    + 2 |{\d\d\u}\rangle \bigr] \notag \\[2ex]
   |\Psi_3\rangle &\equiv& |\tfrac{1}{2},\tfrac{+1}{2}\rangle^- =
   \tfrac{1}{\sqrt{2}} \bigl|(\u\d {-} \d\u)\u\bigr\rangle \notag \\
   |\Psi_4\rangle &\equiv& |\tfrac{1}{2},\tfrac{-1}{2}\rangle^- =
   \tfrac{1}{\sqrt{2}}\bigl|(\d\u {-} \u\d)\d\bigr\rangle
   \notag \\[2ex]
   |\Psi_5\rangle &\equiv& |\tfrac{3}{2},\tfrac{+3}{2}\rangle = 
    |{\u\u\u}\rangle \notag \\
   |\Psi_6\rangle &\equiv& |\tfrac{3}{2},\tfrac{+1}{2}\rangle = 
   \tfrac{1}{\sqrt{3}}\, \bigl| \u\u\d {+} \u\d\u {+} \d\u\u\bigr\rangle \notag \\
   |\Psi_7\rangle &\equiv& |\tfrac{3}{2},\tfrac{-1}{2}\rangle = 
   \tfrac{1}{\sqrt{3}}\, \bigl| \d\d\u {+} \d\u\d {+} \u\d\d \bigr\rangle \notag \\
   |\Psi_8\rangle &\equiv& |\tfrac{3}{2},\tfrac{-3}{2}\rangle = 
    |{\d\d\d}\rangle % \text{ ,}
\label{eq:basis}
\end{eqnarray}
where $|\sfrac{1}{2}\rangle^\pm$ are the low-energy
doubly degenerate $S=\sfrac{1}{2}$ multiplets with the
eigenvalue of $-\Delta_0/2 \pm \alpha J_1/2$, with
$\Delta_0 \equiv \tfrac{1}{2}(J_2 + 2J_3)$. They are
symmetric ($+$) or antisymmetric ($-$) under rung
exchange, i.e., exchange of sites $m=1,2$. They merge
into a four-fold degeneracy at $\alpha=0$. Otherwise,
there exists an ``orbital'' splitting of $\alpha J_1
\equiv J_2{-}J_3$. The remaining four states are the
eigenvectors that form the high-energy $S=\sfrac{3}{2}$
multiplet with eigenvalue $+\Delta_0/2$ which are
symmetric under rung exchange.

The excitation energy from the low-energy states to
the high-energy states is $\Delta(\alpha)=\Delta_0 \pm
\alpha J_1/2$. For the low-enough temperature $T$
satisfying $e^{-\Delta/T}\ll 1$ (i.e., vanishing
thermal population of the four high-energy states) and
$\Delta/J \gg 1$ (i.e., little quantum fluctuations
between these two groups), the high-energy states are
irrelevant to the low-energy physics. Therefore we
project the Hamiltonian into the space formed by the
four low-energy states using the many-body downfolding
method \cite{Hubbard_X_Operator, White_NCT_02,
Yin_PRB_09_cuprates, Yin_downfolding_conf,
Yin_PRL_LaMnO3, Yin_PRL_Sr3CuIrO6, Yin_PRL_pyroxene}
based on Hubbard operators \cite{Hubbard_X_Operator}.
For the Hamiltonian with spin only operators,
it is convenient to use the following matrix
representation~\cite{Yin_PRL_Sr3CuIrO6}.
The eigenvectors in \Eq{eq:basis} constitute the unitary
transformation (also indicating the order of states to
the left),
\begin{eqnarray}
U = 
\begin{array}[c]{c}
 \u\u\u \\[.5ex]
 \u\u\d \\[.5ex]
 \u\d\u \\[.5ex]
 \u\d\d \\[.5ex]
 \d\u\u \\[.5ex]
 \d\u\d \\[.5ex]
 \d\d\u \\[.5ex]
 \d\d\d 
\end{array}
\begin{pmatrix}
 0 & 0 & 0 & 0 & 1 & 0 & 0 & 0 \\
 \tfrac{-2}{\sqrt{6}} & 0 & 0 & 0 & 0 & \tfrac{1}{\sqrt{3}} & 0 & 0 \\
 \tfrac{1}{\sqrt{6}} & 0 & \tfrac{1}{\sqrt{2}} & 0 & 0 & \tfrac{1}{\sqrt{3}} & 0 & 0 \\
 0 & \tfrac{-1}{\sqrt{6}} & 0 & \tfrac{1}{\sqrt{2}} & 0 & 0 & \tfrac{1}{\sqrt{3}} & 0 \\
 \tfrac{1}{\sqrt{6}} & 0 & \tfrac{-1}{\sqrt{2}} & 0 & 0 & \tfrac{1}{\sqrt{3}} & 0 & 0 \\
 0 & \tfrac{-1}{\sqrt{6}} & 0 & \tfrac{-1}{\sqrt{2}} & 0 & 0 & \tfrac{1}{\sqrt{3}} & 0 \\
 0 & \tfrac{2}{\sqrt{6}} & 0 & 0 & 0 & 0 & \tfrac{1}{\sqrt{3}} & 0 \\
 0 & 0 & 0 & 0 & 0 & 0 & 0 & 1 \\
\end{pmatrix}
\text{. }\qquad
\label{U}
\end{eqnarray}
The projection for any operator $\hat{O}$ is done in
the following procedure: Perform $ U^{T} \hat{O} U $
and retain the entries in the low-lying
$4$-dimensional Hilbert space as the zeroth-order
approximation and/or use the canonical transformation
to get the higher-order terms \cite{White_NCT_02,
Yin_PRB_09_cuprates, Yin_downfolding_conf,
Yin_PRL_LaMnO3, Yin_PRL_Sr3CuIrO6, Yin_PRL_pyroxene}.
The resulting $4\times4$ matrices in the low-energy
regime can be conveniently described by introducing
two auxiliary spin $S=\sfrac{1}{2}$ operators 
\begin{eqnarray}
   \Sspin^a &=& \tfrac{1}{2} \sigma^a \otimes 1^{(2)}
\label{s} \\
   \Torb^a &=& 1^{(2)} \otimes \tfrac{1}{2}\tau^a
\label{tau}
\end{eqnarray}
with $\sigma^a$ and $\tau^a$ the Pauli matrices,
having $a\in \{{x,y,z}\}$, and $1^{(2)}$ the
$2\times2$ identity matrix. Assuming that $\sigma$
represents the fast index in $\sigma \otimes \tau$
(aka., column major ordering), then given the state
ordering in \Eq{eq:basis},
the $\Sspin$ operators are spin-like because they
operate within $|\Psi_1\rangle$ and $|\Psi_2\rangle$,
or within $|\Psi_3\rangle$ and $|\Psi_4\rangle$,
referred to as orbital 1 or 2, respectively.
Conversely, the $\Torb$ operators connect these two
``orbitals'' split by the energy $\alpha J$. Then,
any projected operator $\hat{O}$ can be written in the
basis of the $\Sspin$ and $\Torb$ operators,
\begin{eqnarray}
  \hat{O}_\mathrm{projected} &=& U^{T} \hat{O} U \notag \\
&=& f(I,\Sspin^x,\Sspin^y,\Sspin^z,\Torb^x,\Torb^y,\Torb^z),
  \label{projection}
\end{eqnarray}
where $I$ is the $4\times 4$ identity matrix.

Since the strengths of the zero- and first-order terms
are proportional to $J$ and $J^2/\Delta$,
respectively, it suffices for $J/\Delta\ll 1$ to study
the zeroth order, i.e., the plain projection into the
low-energy regime ~\cite{Yin_PRL_Sr3CuIrO6}.
The projected \emph{inter-rung} interaction $J$
terms in the zeroth-order approximation are given in
\Eq{eq:Heff}. They can be obtained by using the
projected spin operators in the zeroth-order
approximation
\begin{eqnarray}
   S^a_{1} &=& \tfrac{2}{3}\,\Sspin^a\,\bigl(\tfrac{1}{2}I+\Torb^z + \sqrt{3} \Torb^x \bigr), \nonumber\\
   S^a_{2} &=& \tfrac{2}{3}\,\Sspin^a\,\bigl(\tfrac{1}{2}I+\Torb^z - \sqrt{3} \Torb^x \bigr), \label{eq:Snew} \\
   S^a_{3} &=& \tfrac{2}{3}\,\Sspin^a\,\bigl(\tfrac{1}{2}I - 2\Torb^z \bigr), \nonumber
\end{eqnarray}
with $a \in \{x,y,z\}$. The spin operators $\Sspin^a$
have the simple interpretation, that they exactly
represent the total rung spin, i.e., $\Sspin \equiv
\Ssite{1}+\Ssite{2}+\Ssite{3} \equiv {\bf
S}^{\mathrm{tot}}_\mathrm{rung}$. With the low-energy
space fully residing within the $S=\sfrac{1}{2}$
symmetry sector, this is a well-defined spin-half
operator, indeed. However, we stress that for the
projection of the \emph{intra-rung} and general
physical quantities, one should not use \Eq{eq:Snew}.
The correct way is to follow \Eqs{U} -
\eqref{projection}, i.e., first do the exact
transformation in the $8\times 8$ space and then do
the reduction as the very last step. For example, in
the correct way, $S^2_{i,1}=\tfrac{3}{4}$ is correctly
reproduced in both the $8\times8$ and $4\times4$
matrix representations. In contrast, $S^2_{i,1} =
\tfrac{5}{12}I + \tfrac{1}{3}\Torb^z +
\tfrac{1}{\sqrt{3}}\Torb^x$ in the said incorrect way.
This is a consequence of the special algebra of
Hubbard operators for on-site or intra-rung
actions~\cite{Hubbard_X_Operator}.

\section{Mean field treatment with translational invariance enforced}
\label{app:meanfield}

\begin{figure}[t]
\includegraphics[width=0.7\columnwidth,clip=true,angle=0]{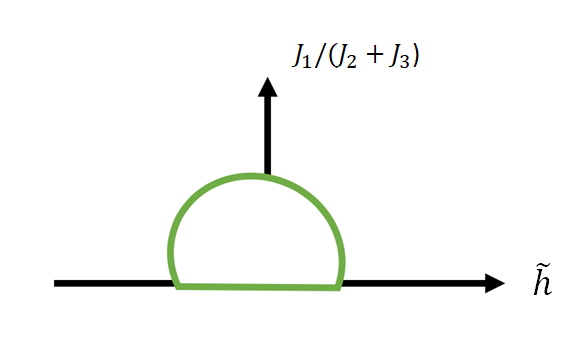}
\caption{
  Schematic phase diagram suggested by the SMF
  analysis vs. $J_1/(J_2{+}J_3)$ and effective orbital
  field $\tilde{h} \sim \alpha \sim J_2 {-} J_3$. The
  leg spins are ferromagnetically ($\langle\Torb_i^z\rangle>0)$
  and antiferromagnetically ($\langle\Torb_i^z\rangle<0$)
  aligned in the two phases $\alpha \ll -1$ and
  $\alpha \gg +1$, respectively. They are separated by
  a phase with spontaneously broken \Ztwo symmetry
  where $\langle\Torb^x\rangle \neq 0$.
}
\label{Fig:QCP}
\end{figure}

Here we show that a semi-mean-field (SMF) treatment
assuming a uniform, i.e., non-dimerized state permits
an entire intermediate phase with
$\langle\Torb^x\rangle \neq 0$ instead of a QCP at
sufficiently small $\alpha$, as schematically depicted
in \Fig{Fig:QCP}. We stress, however, that eventually
this is {\it not realized} in the many-body low-energy
regime of the system, in that DMRG clearly finds a
gapped dimerized ground state, instead. Nevertheless,
we believe this still represents an interesting point
of view, hence we present this here in the appendix.
To start with the SMF treatment, we assume
translational invariance, and perform a mean-field
decoupling of the spin from the orbital degrees of
freedom [cf. \Eqs{eq:Heff}],
\begin{eqnarray}
  \tilde{\mathcal{H}} &=&
    \tilde{\mathcal{H}}_\mathrm{spin}
  + \tilde{\mathcal{H}}_\mathrm{orb}
  - \tilde{\mathcal{E}}_0
\label{eq:meanfield}
\end{eqnarray}
where
\begin{subequations} 
\begin{eqnarray}
  \tilde{\mathcal{H}}_\mathrm{spin} &\equiv&
  \sum_i
  \tilde{J}_{\Sspin} 
   \mathbf{\Sspin}_i \cdot \mathbf{\Sspin}_{i+1} 
\label{eq:meanfield:spin} \\
  \tilde{\mathcal{H}}_\mathrm{orb} &\equiv&
  \sum_i \Bigl(
  - \tilde{J}_{\Torb} \bigl(\Torb_i^z\Torb_{i+1}^z
  + 3 \Torb_i^x\Torb_{i+1}^x \bigr)
  + \tilde{h} \Torb_i^z \Bigr)
\label{eq:meanfield:orb} \\
  \tilde{\mathcal{E}}_0 &\equiv&
   \tfrac{8 J_1}{9} N
   \langle \mathbf{\Sspin}_i {\cdot}\mathbf{\Sspin}_{i+1}\rangle
   \langle\Torb_i^z+\Torb_i^z\Torb_{i+1}^z+3\Torb_i^x\Torb_{i+1}^x \rangle
\label{eq:meanfield:E0}
% \text{, }
\ \qquad
\end{eqnarray}
\end{subequations} 
with the effective mean-field couplings
\begin{subequations} 
\begin{eqnarray}
  \tilde{J}_{\Sspin} &\equiv& \tfrac{8 J_1}{9} \bigl\langle 
       \tfrac{1}{4}
     + \Torb_i^z 
     + \Torb_i^z\Torb_{i+1}^z + 3 \Torb_i^x\Torb_{i+1}^x 
  \bigr\rangle \ge 0  \qquad
\label{eq:Js:MF}
\\
  \tilde{J}_{\Torb} &\equiv& - \tfrac{8 J_1}{9} \langle
      \mathbf{\Sspin}_i\cdot\mathbf{\Sspin}_{i+1}
   \rangle = \alpha_0 J_1 >0 
\label{eq:Jt:MF}
\\
  \tilde{h} \ &\equiv& J_1 \Bigl(
     \alpha +\underbrace{\tfrac{8}{9} \langle
        \mathbf{\Sspin}_i\cdot \mathbf{\Sspin}_{i+1}
     \rangle
     }_{\equiv -\alpha_0}
  \Bigr)
\text{ .}\label{eq:heff}
\end{eqnarray}
\end{subequations}
Here \Eq{eq:meanfield:E0} is just the mean-field
reference energy, with the various local expectation
values assumed independent of $i=1,\ldots,N$.
The decoupled spin and orbital sectors,
\Eq{eq:meanfield:spin} and \Eq{eq:meanfield:orb},
respectively, can be solved self-consistently now
given their respective quantum Hamiltonians (hence the
terminology `semi-mean-field'). Having assumed
translational invariance, the spin Hamiltonian
\eqref{eq:meanfield:spin} is always gapless. In
contrast, for large $|\tilde{h}|$, the orbital
Hamiltonian \eqref{eq:meanfield:orb} is always gapped.
Its ground state determines the active orbital in the
spin Hamiltonian \eqref{eq:meanfield:spin}.

The resulting schematic SMF phase diagram, assuming a
non-dimerized phase, is depicted in \Fig{Fig:QCP}. We
shall briefly discuss its three phases.
For the ground state of a spin-half Heisenberg chain
one has the exact result, $\langle
\mathbf{\Sspin}_i\cdot\mathbf{\Sspin}_{i+1}\rangle =
\tfrac{1}{4}-\ln{2}$ \cite{Hulthen_Heisenberg_GS,
Bethe_ansatz_1931}. Therefore assuming that the spin
sector is close to its ground state, one obtains
$\alpha_0 \approx 0.394$ for $T=0$ [cf. \Eq{eq:heff}].
For the value $\alpha=\alpha_0$ then, i.e., $\tilde{h}=0$,
the dominance of the symmetric or antisymmetric
$S=\sfrac{1}{2}$ rung multiplet switch roles.

In the orbital sector, the Hamiltonian
\eqref{eq:meanfield:orb} has quantum critical points
in the same universality class as the quantum Ising
model with ferromagnetic interaction. The interaction
strength for the $\Torb_i^x\Torb_{i+1}^x$ term is
three times as large as that for the
$\Torb_i^z\Torb_{i+1}^z$ term. In the continuum limit the
$\Torb^z$ operator becomes a product of right and left
moving Majorana fermions and hence the term
$\Torb^z_i\Torb^z_{i+1} \sim \rho_i \chi_{i}
\rho_{i+1}\chi_{i+1} \sim \rho\partial_x\rho
\chi\partial_x\chi$ [cf. \Eq{JW} in the main text]
becomes highly irrelevant with a scaling dimension of
$d=4$. In the absence of the $\Torb^z \Torb^z$ term, the
criterion for the emergence of the symmetry broken
state with finite $\langle \Torb^z \rangle$ can be
estimated by \cite{Pfeuty_Ising_field} 
\begin{equation}
   \bigl|\tfrac{2\tilde{h}}{3 \tilde{J}_{\Torb}}\bigr|
  = \tfrac{2}{3}\, \bigl|\tfrac{\alpha-\alpha_0}{\alpha_0}\bigr|
  < 1, 
\end{equation}
given the critical field $|\tilde{h}|_\mathrm{cr}
\simeq \tfrac{(3\tilde{J}_T)}{2}$ [cf.
\Eq{eq:meanfield:orb}]. This corresponds to $\alpha
\in \tfrac{\alpha_0}{2}\, [-1,5] \approx [-0.197,
0.985]$. The neglected $\Torb^z \Torb^z$ term is
expected to shift these boundaries, as motivated by a
mean-field decoupling $\Torb^z \langle\Torb^z\rangle$. 

Right at $\tilde{h}=0$, the orbital Hamiltonian
becomes a version of the XY model in zero magnetic field
where exact results for the magnetization are
available: $\langle \Torb_i^x\rangle = 
\tfrac{\sqrt{2}}{3}\approx
0.471$~\cite{Lieb_XY_Ising_field_exact}. The state
with a spontaneously broken \Ztwo symmetry can be
understood as the state where the center spins
predominantly form singlets with a particular leg of
the ladder which would translate into an asymmetry of
correlation functions that include leg spins. When
$\alpha$ increases, the system undergoes a phase
transition into the symmetric state with nonzero
$\langle\Torb^z_i\rangle$ where the above asymmetry
disappears.
For any finite temperature $T$, the symmetry is
restored by thermal average, i.e., having $\langle
\Torb_i^x\rangle=0$, whereas $\langle
\Torb_i^z\rangle$ is proportional to the effective
field when it is weak \cite{Pfeuty_Ising_field,
Lieb_XY_Ising_field_exact}. As a result, it does not
contain a phase transition at finite temperature.

For large $\tilde{h}$ the orbitals become strongly
polarized, as discussed in the main text. With
$\langle\Torb^z_i\rangle \simeq \pm \tfrac{1}{2}$,
the effective spin coupling in \Eq{eq:Js:MF} becomes
$\tilde{J}_{\Sspin} \simeq \tfrac{8 J_1}{9}
(\tfrac{2}{4} \pm \tfrac{1}{2})$, which thus motivates
the positive sign indicated with \Eq{eq:Js:MF}.
For example, for dominant $J_2$, i.e., $\alpha \gg +1$
with $\langle\Torb^z_i\rangle \simeq -\tfrac{1}{2}$,
the center spins become nearly decoupled, which thus
corresponds to a spin-half Heisenberg chain with
vanishing effective coupling $\tilde{J}_{\Sspin}\sim
0$ in the low-energy regime of the system. Conversely,
for dominant $J_3$, i.e., $\alpha \ll -1$ with
$\langle\Torb^z_i\rangle \simeq +\tfrac{1}{2}$,
the low energy behavior is described by a single
effective Heisenberg chain with finite effective
coupling $\tilde{J}_{\Sspin} \simeq \tfrac{8J}{9}$.
Note that the same picture for large $\tilde{h}$,
and hence large $\alpha$, already also applies in the
original Hamiltonian \eqref{eq:Heff} in the main
text, and hence is not constrained to the mean-field
analysis here.

\section{Absence of \Ztwo symmetry breaking in DMRG}
\label{app:absenceZ2}

\begin{figure}[t]
\begin{center}
\includegraphics[width=1\linewidth]{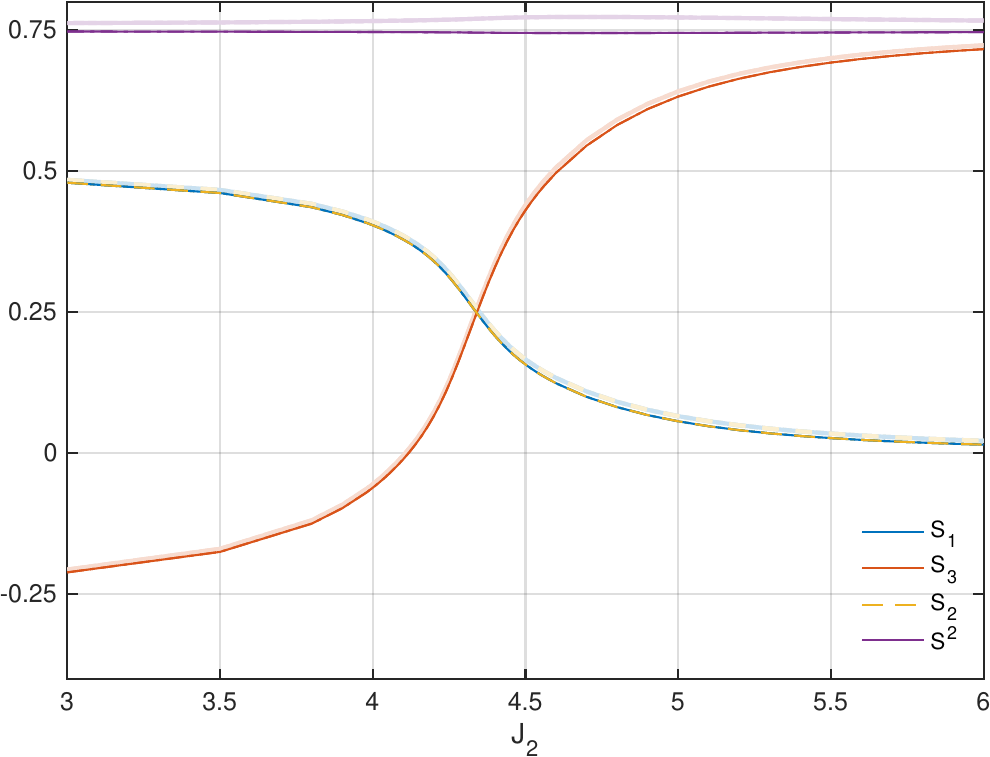}
\end{center}
\caption{
  Local DMRG expectation values as in \Eqs{eq:Si:loc}
  with focus on the presence of a nematic phase
  for the full Hamiltonian \eqref{eq:model}
  for $L=64$ and $J=[1,J_2,4]$, as also analyzed
  in \Fig{fig:dim_gap}, having $S^2 \equiv S_1 + S_2 + S_3$.
  Light colors use the full
  local spin operators $\Ssite{m}$, whereas strong
  colors use the projected spin operators ${\bf S}_m$,
  as reflected in \Eqs{eq:Si:loc}.
  The ground state is the same in either case,
  hence also includes the local $S=\sfrac{3}{2}$
  multiplet.
}
\label{fig:tmp_Si}
%\end{figure}
%
%\begin{figure}[tbh!]
\vspace{.25in}
\includegraphics[width=1\columnwidth,clip=true,angle=0]{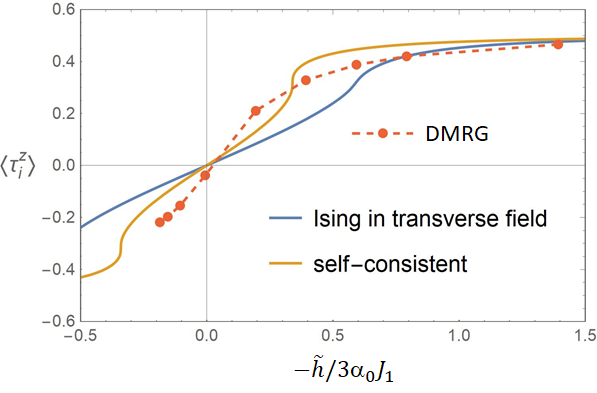}
\caption{
   Comparison of the orbital expectation value
   $\langle\Torb_i^z\rangle$ vs. $\tilde{h}$ between
   DMRG and SMF: DMRG data as in \Fig{fig:tmp_Si},
   the exact mean-field solution of \Eq{eq:Js:MF}
   in the absence of the $\langle\Torb_i^z\rangle$ term
   (i.e., the Ising model in a transverse
   field~\cite{Pfeuty_Ising_field}), and the
   self-consistent mean-field theory for \Eq{eq:Js:MF}.
}
\label{Fig:Tz}
\end{figure}

In the strong coupling limit we have [cf. \Eqs{eq:Snew}]
\begin{eqnarray}
  S_{m}
  &\equiv& \langle {\bf S}_m\! \cdot\! \Sspin \rangle_i
  = \tfrac{2 s^2}{3}\bigl(
    \tfrac{1}{2}
   +\langle\Torb_i^z\rangle \pm \sqrt 3\langle\Torb_i^x\rangle
 \bigr), \ \  (m=1,2)
\notag \\
  S_{3}
  &\equiv & \langle {\bf S}_3 \cdot \Sspin \rangle_i
  = \tfrac{2 s^2}{3} \bigl(
    \tfrac{1}{2} -2\langle\Torb_i^z\rangle\bigr)
,\label{eq:Si:loc}
\end{eqnarray}
where $s^2$ is the spectral weight in the spin sector.
Therefore $S_1$ can be negative only in the nematic
phase, whereas $S_3 <0$ is permitted more generically,
namely when $\langle\Torb^z\rangle > 1/4$.

We evaluated the expectation values in \Eq{eq:Si:loc}
using DMRG, with the results summarized in
\Fig{fig:tmp_Si}. Given that $S_1$ and $S_2$ stay
positive, there is clearly no support for a nematic
phase. Besides, the data for $S_1$ and $S_2$ lies
exactly on top of each other, which thus also
demonstrates that the rung exchange symmetry is
preserved. The local expectation value $S_3$ can
become negative, but that simply reflects orbital
polarization.
As already seen with \Fig{fig:dim_gap} in the main
text, dimerization is only visible for expectation
values that stretch along the system. For expectation
values within individual rungs, this data is the same
for even and odd rungs, i.e., does not display
dimerization in itself. This also holds in the
present case for the data in \Fig{fig:tmp_Si}.
 
Adding up the data, $S_1 + S_2 + S_3 = S^2$, this
yields the expectation value of the total spin
operator (also labeled $S^2$ in \Fig{fig:tmp_Si})
which is approximately constant, having $S^2\approx
0.75$. This demonstrates that the present parameter
setting with $J_3=4$ is deep within the strong
rung-coupling regime, in that the local density matrix
is overwhelmingly dominated by the $S=\sfrac{1}{2}$
multiplets. The $S^2$ data in light color reaches
slightly above $0.75$ which shows that it also
includes a weak $S=\sfrac{3}{2}$ component. The $S^2$
data in strong color is slightly deficient of $0.75$,
because it refers to the projected spin operators.

Overall, our DMRG data here again finds no evidence
for the {\Ztwo}-symmetry-broken phase with
$\langle\Torb^x\rangle \neq 0$ near $\tilde{h}=0$, as
suggested by a semi-mean-field analysis on a uniform
system. Instead, bond dimerization is found. To
evaluate what has been missed in the SMF analysis
assuming the translational invariance, we compare
the obtained $\langle\Torb_i^z\rangle$ with the DMRG
result, as shown in Fig.~\ref{Fig:Tz}. The vertical
slopes that indicate the phase boundaries of the
intermediate \Ztwo-broken phase in the
self-consistent SMF analysis, are entirely absent in
the DMRG data which evolves smoothly, throughout.
The considerably stronger $\langle\Torb_i^z\rangle$
values in the DMRG data near $\tilde{h}=0$ suggests
that the semi-mean-field theory needs to allow for
bond dimerization, as discussed in the main text.

\end{document}